\documentclass[12pt]{article}
\usepackage[dvipdfmx]{graphicx}
\usepackage[margin=30truemm]{geometry}
\usepackage{amsmath}
\usepackage{amssymb}
\usepackage{authblk}
\usepackage{mathrsfs}
\usepackage{slashed}
\usepackage{color}
\usepackage{cite}
\usepackage{here}
\usepackage{subcaption}
\usepackage[normalem]{ulem}
\usepackage{wrapfig}

\makeatletter
 
 \@addtoreset{equation}{section}
 \makeatother

\newcommand{\be}{\begin{equation}}
\newcommand{\ee}{\end{equation}}

\newcommand{\bea}{\begin{equnarray}}
\newcommand{\eea}{\end{eqnarray}}
\newcommand{\beann}{\begin{equnarray*}}

\newcommand{\eeann}{\end{eqnarray*}}
\newcommand{\nn}{\nonumber}
\newcommand{\ba}{\begin{array}}
\newcommand{\ea}{\end{array}}

\newcommand{\bs}{\boldsymbol}

\newcommand{\bsa}{{\boldsymbol{a}}}
\newcommand{\bsb}{{\boldsymbol{b}}}
\newcommand{\bse}{{\boldsymbol{e}}}
\newcommand{\bsf}{{\boldsymbol{f}}}
\newcommand{\bsone}{\bs{1}}

\newcommand\dashint{\mathchoice
  {\int\kern-10pt-}
  {\int\kern-8.5pt-}
  {\int\kern-6.1pt-}
  {\int\kern-4.58pt-}}

\DeclareMathOperator{\diag}{diag}

\DeclareMathOperator{\Tr}{Tr}

\DeclareMathOperator{\Real}{Re}

\title{Phases and Duality\\in Fundamental Kazakov-Migdal Model\\on the Graph}
\author[1]{So Matsuura\thanks{s.matsu@keio.jp}}
\author[2]{Kazutoshi Ohta\thanks{kazutoshi.ohta@mi.meijigakuin.ac.jp}}
\affil[1]{\it Hiyoshi Departments of Physics,
and Research and Education Center for Natural Sciences,
Keio University, 
Yokohama, Kanagawa 223-8521, Japan}
\affil[2]{\it Institute for Mathematical Informatics,
Meiji Gakuin University, Yokohama, Kanagawa 244-8539, Japan}

\date{}

\begin{document}
\maketitle

\begin{center}
{\bf Abstract}
\end{center}

We examine the fundamental Kazakov-Migdal (FKM) model on a generic graph,
whose partition function is represented by the Ihara zeta function weighted by unitary matrices. 
The FKM model becomes unstable in the critical strip of the Ihara zeta function.
We discover a duality between small and large couplings, associated with the functional equation of the Ihara zeta function for regular graphs.
Although the duality is not precise for irregular graphs, we show that the effective action in the large coupling region can be represented by a summation of all possible Wilson loops on the graph similar to that in the small coupling region.
We estimate the phase structure of the FKM model both in the small and large coupling regions by comparing it with the Gross-Witten-Wadia (GWW) model.
We further validate the theoretical analysis through detailed numerical simulations.

\newpage

\section{Introduction}
\label{sec:Introduction}

Gauge theory serves as a fundamental framework for describing the interactions in nature.
Specifically, the property of the asymptotic freedom in the non-Abelian gauge theory, which describes QCD, 
imposes limitations on analysis solely through perturbation theory.
As a result, non-perturbative approaches, especially lattice gauge theories \cite{PhysRevD.10.2445}, have emerged as crucial tools. 
A characteristic feature of lattice gauge theory is that it divides continuous space-time into discrete lattices (graphs) and approximates infinite degrees of freedom with finite degrees of freedom.
This not only non-perturbatively defines the theory but also enables the application of powerful numerical analysis techniques.

In our recent series of studies \cite{matsuura2022kazakov, matsuura2022graph},
we have developed a discretized gauge theory on graphs as a generalization
of the so-called Kazakov-Migdal (KM) model \cite{kazakov1993induced}. 
The KM model on the graph has a mathematically remarkable property that its partition function is represented by the graph zeta functions, that is, graph-theoretical analogues of the Riemann zeta function  
\cite{Ihara:original,MR607504,sunada1986functions,Hashimoto1990ONZA,bass1992ihara,bartholdi2000counting,mizuno2003bartholdi}
(For comprehensive reviews and references, see also \cite{terras_2010}).
This enables the generalized KM model to accurately enumerate Wilson loops on the graph and to precisely evaluate the partition function in the large $N_c$ limit.
While the generalized KM model displays fascinating characteristics, its application as a model for induced QCD — the foundational motivation behind the KM model — faces the same difficulties as the original KM model due to the presence of an undesirable local $U(1)$ symmetry \cite{kogan1992induced,kogan1993area,kogan1993continuum, migdal1993bose,cline1993induced,balakrishna1994difficulties}.

To construct a model closer to QCD, 
we have replaced the scalar fields of the generalized KM model in the adjoint representation with ones in the fundamental representation.
We call this model the fundamental Kazakov-Migdal (FKM) model \cite{PhysRevD.108.054504},
as a generalization of the model introduced in \cite{Arefeva:1993ik}.
The FKM model does not have the extra $U(1)$ symmetry with retaining the property that the partition function is represented by the graph zeta function. 
As a result, the effective action of the FKM model includes the conventional Wilson action of gauge theory as a limit of certain theoretical parameters, suggesting the model's potential to describe a realistic gauge theory.
We have analytically shown that the FKM model on the cycle graph undergoes the Gross-Witten-Wadia (GWW) third-order phase transition
\cite{Gross:1980he,Wadia:1980cp}, 
and have found theoretical evidences for the existence of non-trivial phase transitions for the general graphs, although the order of these phase transition is not so clear.

In this paper, we investigate the phase structure of the FKM model whose partition function is described by the Ihara zeta function, which is the ancestor of the graph zeta function, in more detail.
Further theoretical and numerical analysis yields the following interesting results on the phase structure of the FKM model:

First of all, the FKM model on a regular graph, where each vertex has the same degree, exhibits a duality between the small and large coupling regions.
This duality is a consequence of the functional equation of the Ihara zeta function on the regular graph. 
Secondly, we discover a dual transformation applicable to the Ihara zeta function of any graph, including irregular graphs, as a generalization of the functional equation of the Ihara zeta function of the regular graph. 
While the exact duality does not hold for the FKM model on irregular graphs, the effective action in the large coupling region still can be determined by using the dual transformation.
In particular, the phase structure of the FKM model, even in the large coupling region, can be elucidated through the asymptotic expansion of the effective action, especially when the number of ``flavors'' is large enough for the GWW model to become a good approximation of the FKM model.
Analytical and numerical investigations of the phase structure of the FKM model on various graphs reveals that it is closely related to the structure of the graph.

The organization of this paper is as follows:
 
In the next section, 
we introduce the notation of the graph theory and the Ihara graph zeta function used in this paper, and then briefly review the basic properties of the FKM model.
In Sec.~\ref{sec:FuncEq}, we explain the functional equation of the Ihara zeta function, which is the origin of the duality of the FKM model. 
In Sec.~\ref{sec:duality}, 
we show that there is an unstable parameter region in the FKM model corresponding to the distribution of the poles of the Ihara zeta function and explain the duality of the FKM model.
After discussing the exact results of the cycle graph, 
we theoretically analyze the phase structure of the FKM model on some specific examples of both regular and irregular graphs.
In Sec.~\ref{sec:Regular}, we give the results of the numerical simulation for the regular graphs.
The exact analytical results for the cycle graph
is used for the check of the simulation code. 
In Sec.~\ref{sec:Irregular}, we show the results of the numerical simulation for the irregular graphs. 
The last section is devoted to the conclusion and discussions.

\section{Review of the Fundamental Kazakov-Migdal Model}
\label{sec:FKM}

\subsection{Notations in the graph theory}

The graph $G$ is an object which consists of
vertices and edges that connect two vertices. 
We denote the set of the vertices and edges by $V$ and
$E$, respectively,
with their cardinalities represented as $n_V$ and $n_E$. 
We assume that the edges have directions. 
Therefore, the term ``graph'' within this paper refers to a directed graph (digraph). 
A directed edge $e$ starting from a vertex $v$ and ending at a vertex $v'$ is represented as $\langle v,v'\rangle$. 
The two vertices $v$ and $v'$ of $e=\langle v,v'\rangle$ are called 
the ``source'' $v=s(e)$ and the ``target'' $v'=t(e)$ of $e$, respectively.
The degree of a vertex $v$, denoted as $\deg v$, is the number of its neighboring vertices. 
It is useful to consider 
both a directed edge $e=\langle v,v'\rangle$ and its opposite $e^{-1}=\langle v',v\rangle$. 
We then enhance the set of the edges $E$ to
$E_D=\{\bse_a|a=1,\cdots,2n_E\}
=\{e_1,\cdots, e_{n_E}, e_1^{-1}, \cdots e_{n_E}^{-1}\}$,
combining with the opposite directed edges.
In the following, we assume that there is at most one edge connecting two vertices, and  no edge connecting the same vertex. In other words, we consider only simple digraphs in this paper.

A path (or walk) of length $l$ on the graph $G$ is a sequence of $l$ edges, $P=\bse_{a_1} \bse_{a_2} \cdots \bse_{a_l}$ satisfying the conditions $t(\bse_{a_i})=s(\bse_{a_{i+1}})$ ($i=1,\cdots,l$ with $a_{l+1}=a_1$). 
The inverse of the path $P$ is defined as $P^{-1}=\bse_{a_l}^{-1}\cdots \bse_{a_1}^{-1}$. 
A path $C$ whose starting and ending vertices are the same is called a cycle.
We can define the product of two paths $P_1=\bse_{a_1} \cdots \bse_{a_{l_1}}$ and $P=\bse_{a'_1} \cdots \bse_{a'_{l_2}}$ when $t(\bse_{a_l})=s(\bse_{a'_1})$ as $P_1P_2 =\bse_{a_1} \cdots \bse_{a_{l_1}}\bse_{a'_1} \cdots \bse_{a'_{l_2}}$. 
In particular, a power of a cycle $C^n$, which is constructed through this product rule, is also a cycle. 
In this paper, we consider only connected graphs, that is, graphs where any pair of vertices is connected by a path. 
We also assume that the degrees of all vertices are more than or equal to two and the graph is simple, that is, it does not contain more than one edge between two vertices 
and does not contain any edge that connects a vertex to itself (loop).

A part of a path of length $l$ satisfying $\bse_{a_{j+1}}=\bse_{a_{j}}^{-1}$ ($j=1,\ldots, l-1$) is called a backtracking. 
If a cycle of length $l$ satisfies $\bse_{a_l}=\bse_{a_1}^{-1}$, this part is called a tail of the cycle $C$. 
For a cycle $C$, the equivalence class $[C]$ is defined by the set of cyclic rotations of $C$ as 
\[[C] = \{\bse_{a_1} \bse_{a_2} \cdots \bse_{a_l}, \,
\bse_{a_2} \cdots \bse_{a_l}\bse_{a_1}, \, \ldots \,,
\bse_{a_l}\bse_{a_1}\cdots\bse_{a_{l-1}}\}\,. \]
For the equivalence class of the cycle $[C]$, the backtracking and the tail are identical, hence they are collectively called a bump.

A cycle $C$ is called reduced when it contains no bump. 
A cycle $C$ is called primitive when $C$ does not satisfy $C \ne B^r$ for any cycle $B$ and $r\ge 2$.
We denote the set of representatives of reduced cycles as $[{\mathcal P}_R] \subset [{\mathcal P}]$.
Since a primitive reduced cycle $C$ has its inverse $C^{-1}$ also be a primitive reduced cycle of equal length, 
the set $[{\mathcal P}_R]$ can be partitioned into two disjoint unions;  
$[{\cal P}_R]=[\Pi_{+}] \sqcup [\Pi_{-}]$, 
where $[\Pi_{-}]$ consists of the inverses of elements in $[\Pi_{+}]$. These elements in $[\Pi_{+}]$ are referred to as (the representatives of) chiral primitive reduced cycles.

\subsection{Ihara zeta functions}

The Ihara zeta function of a graph $G$ is defined by
a product over the primitive reduced cycles on $G$ \cite{Ihara:original,MR607504,sunada1986functions}
\begin{equation}
  \zeta_G(q)\equiv
  \prod_{C\in [{\cal P}_R]} \frac{1}{1-q^{|C|} }\,,
  \label{eq:Ihara}
\end{equation}
which is an analog to the Euler product expression
of the Riemann zeta function.
In the above expression,
$|C|$ denotes the length of the cycle $C$. 
One of the important properties of the Ihara zeta function is that it is expressed as the reciprocal of a polynomial
\begin{equation}
  \zeta_G(q) = \bigl(1-q^2\bigr)^{-(n_E-n_V)}
  \det\bigl(\bsone_{n_V}-q A + q^2Q\bigr)^{-1}\,,
  \label{eq:vertex Ihara}
\end{equation}
where 
a square matrix $A$ of size $n_V$ is called the adjacency matrix
defined by
\begin{equation}
A_{vv'} = \sum_{\bse\in E_D}
\delta_{\langle{v},{v'}\rangle,\bse}   \quad (v,v'\in V)\, ,
\label{eq:matrix A}
\end{equation}
and $Q$ is a diagonal matrix given by
\begin{equation}
Q 
\equiv \diag\left(t_1,t_2,\cdots,t_{n_V}\right) \,,
\label{eq:Q matrix}
\end{equation}
with $t_i \equiv \deg v_i - 1$.
The Ihara zeta function (\ref{eq:Ihara}) has another expression introduced by Hashimoto \cite{Hashimoto1990ONZA,bass1992ihara}, 
\begin{equation}
  \zeta_G(q) = \det \left(\bsone_{2n_E}- qW\right)^{-1}\,,
  \label{eq:edge Ihara}
\end{equation}
where a square matrix $W$ of size $2n_E$ is called the edge adjacency matrix whose elements are given by 
\begin{align}
    W_{\bse\bse'} = \begin{cases}
      1 & {\rm if}\ t(\bse) = s(\bse')\ {\rm and}\ \bse'^{-1}\ne \bse \\
      0 & {\rm others}
    \end{cases}\,,
    \label{eq:matrix W}
\end{align}
where $\bse, \bse' \in E_D$.

\subsection{The FKM model}
On the graph $G$, we can construct a gauge theory by putting
the scalar field $\Phi^I_v$ on the vertices and the gauge fields $U_e$ on the edges.
The edge variables $U_e$ are $N_c\times N_c$ unitary matrices 
and the scalar fields $\Phi^I_v$ transform in the fundamental representation of $U(N_c)$,
where the index $I=1,2,\cdots, N_f$ labels the ``flavors''.
Using these fields, we define a model by the action, 
\be
  S= \sum_{v\in V} m_v^2 \Phi_{v}^{\dagger I} \Phi_{v I}
  -q\sum_{e\in E}\left( \Phi_{s(e)}^{\dagger\ \ I} \, U_e\, \Phi_{t(e) I} +  \Phi_{t(e)}^{\dagger\ \ I}\, U_e^\dagger\, \Phi_{s(e) I} \right) \,,
  \label{eq:action}
\ee
where $q$ is a coupling constant%
\footnote{
Although $q$ is called a coupling constant in this paper, 
we should note that it becomes the inverse of the gauge coupling constant when we compare the model with the GWW model. 
Therefore, the small/large region of $q$ in this paper corresponds to the strong/weak coupling region in the sense of the gauge coupling. 
} 
and $m_v$ are the masses for the scalar fields.
We call this model the fundamental Kazakov-Migdal (FKM) model introduced in \cite{PhysRevD.108.054504}. 
Note that the term proportional to $q$ in the action represents the neighborhood
interaction of $\Phi_v^I$.

In contrast to the traditional Kazakov-Migdal model, where the scalar fields are in the adjoint representation and the effective action of the adjoint scalar fields rather than the edge variables is obtained by using the Harish-Chandra-Itzykson-Zuber integral \cite{kazakov1993induced},
we first integrate out the scalar fields $\Phi_v^I$ in the partition function,
\be
Z_G = \int \prod_{I=1}^{N_f}\prod_{v\in V}d\Phi_{vI}d\Phi^{\dag vI}
\prod_{e\in E}dU_e\,
e^{- S}\ .
\ee
Then, we obtain 
\be
Z_G = 
\left(
2\pi
\right)^{N_cN_fn_V}
\int \prod_{e\in E} dU_e\, 
\det\left(
m_v^2{\delta_v}^{v'}{\bf 1}_{N_c}
-q {(A_U)_v}^{v'}
\right)^{-N_f}\ ,
\label{partition function over U}
\ee
where $A_U$ is a matrix of size $N_c n_V$ 
which is essentially the adjacency matrix \eqref{eq:matrix A} but is weighted by unitary matrices as 
\begin{equation}
{(A_U)_v}^{v'} = \sum_{\bse \in E_D}U_{\bse}
\delta_{\langle v,v'\rangle, \bse}\ .
\label{eq:matrix AU}
\end{equation}
Note that $U_{e^{-1}}$ is regarded as $U_e^{\dag}$ with respect to the
unitary matrix $U_e$ along the original direction.

The integrand of the partition function (\ref{partition function over U})
can be expanded in terms of the various kind of the Wilson loops along
the cycles on $G$. 
This is a reason why this model appears to have the potential to induce QCD \cite{Arefeva:1993ik}, 
but it is difficult to control the coupling constant of the Wilson loops systematically
because of the mismatch between the order of the mass parameter $m_v$ and the net length of the Wilson loop (cycles). 

This difficulty is resolved by tuning the mass parameter $m_v$, following \cite{PhysRevD.108.054504}, as
\be
m_v^2=1+q^2(\deg v-1)=1+q^2 t_v\,. 
\label{coupling tuning}
\ee
Then, the partition function (\ref{partition function over U}) can be rewritten through the unitary matrix weighted Ihara zeta function \cite{matsuura2022kazakov,matsuura2022graph}%
\footnote{$Q$ has been redefined to
$Q=\diag(t_1 \bsone_{N_c}, t_2 \bsone_{N_c}, \cdots, t_{n_V} \bsone_{N_c})$,
but we believe there is no confusion to the matrix $Q$ used to express the original Ihara zeta function \eqref{eq:vertex Ihara}. }, 
\be
  \zeta_G(q;U) = \bigl(1-q^2\bigr)^{-{N_c}(n_E-n_V)}
  \det\bigl(\bsone_{{N_c}n_V}-q A_U + q^2 Q\bigr)^{-1}\,,
  \label{eq:vertex U-Ihara}
\ee
as
\begin{align}
  Z_G &= \left({2\pi}\right)^{{N_f}{N_c}n_V}
  \left(1-q^2\right)^{{N_f}{N_c}(n_E-n_V)}
  \int\prod_{e\in E}dU_e\,
  \zeta_G(q;U)^{N_f}\ .
  \label{eq:Zpre}
\end{align}
Similar to the original Ihara zeta function \eqref{eq:Ihara}, the matrix weighted Ihara zeta function can be expressed in the Euler product form, 
\be
  \zeta_G(q;U) \equiv \prod_{C\in [{\cal P}_R]} \frac{1}{\det\bigl( \bsone_{{N_c}} - q^{|C|}  U_C \bigr)}\ .
  \label{eq:U-Ihara}
\ee
Expanding the product (\ref{eq:U-Ihara}) by the Wilson loops $U_C$, 
we can exactly count the length of the cycles
in the powers of $q$ as 
\begin{align}
  Z_G &= \left({2\pi}\right)^{{N_f}{N_c}n_V}
  \left(1-q^2\right)^{{N_f}{N_c}(n_E-n_V)} \nn \\
  &\qquad \times \int\prod_{e\in E}dU_e\,
  \prod_{C\in [\Pi_{+}]}
  \exp\left\{
    N_f \Tr \left[\sum_{n=1}^\infty \frac{q^{n|C|}}{n}
    \left( 
       U_C^{n} + U_C^{\dagger\, n}  
    \right)
   \right]\right\}\ .
  \label{eq:Z}
\end{align}

Note that we can generalize the parametrization \eqref{coupling tuning} as
\begin{equation}
m_v^2=1-q^2(1-u)^2+q^2(1-u)\deg v\,,
\label{eq:mv Bartholdi}
\end{equation}
so that the partition function is expressed by the matrix weighted Bartholdi zeta function \cite{bartholdi2000counting}, 
which is one parameter extension of the Ihara zeta function. 
However, we focus on the Ihara zeta function in order to simplify the following discussion in this paper. 
The generalization of the analysis of this paper to the Bartholdi zeta function will be discussed separately.

From the partition function (\ref{eq:Z}), the effective action of the model can be read off as 
\begin{align}
    S_{\rm eff}(q;U) 
    &= -N_f \log \zeta_G(q;U) \nn \\
    &=
    -\gamma N_c \Tr\left[\sum_{C\in [\Pi_{+}]}
    \sum_{n=1}^\infty \frac{q^{n|C|}}{n}
    \left( 
      U_C^{n} + U_C^{\dagger\, n}
    \right)\right]\,,
    \label{eq:Seff}
\end{align}
where we have defined $\gamma \equiv N_f/N_c$ and ignored an irrelevant constant term. 
This action is a generalization of the Wilson action of the lattice gauge theory on the graph $G$ in the sense that it contains infinitely many different Wilson loops controlled by the Ihara zeta function. 
So it is interesting to investigate the dynamics and phases of
this complicated model including the higher order interactions. 
 
We note that the unitary matrix weighted Ihara zeta function (\ref{eq:U-Ihara}) also has the Hashimoto expression \cite{Hashimoto1990ONZA} similar to \eqref{eq:edge Ihara},
\begin{align}
\zeta_G(q;U)
&= \det\left( \bsone_{2N_c n_E} - qW_U \right)^{-1}\,,
\label{eq:edge U-Ihara}
\end{align}
where the matrix $W_U$ is an extension of the edge adjacency matrix \eqref{eq:matrix W} weighted by unitary matrices, 
\begin{align}
  (W_U)_{\bse\bse'} = \begin{cases}
  U_{\bse} & {\rm if}\ t(\bse) = s(\bse')\ {\rm and}\ \bse'^{-1}\ne \bse \\
    0 & {\rm others}
  \end{cases}\,. 
  \label{eq:matrix WU}
\end{align}
The equivalence of both expressions (\ref{eq:U-Ihara}) and (\ref{eq:matrix WU})
has been proved in  \cite{matsuura2022graph}
by using the similar technique given by Bass \cite{bass1992ihara}.

\subsection{Gauge fixing and the path representation}
\label{subsec:gauge fixing}

So far, we have considered unitary matrices on all the edges, but we can fix the gauge as $U_e=1$
on a spanning tree (maximal tree) \cite{PhysRevD.15.1128}.
The spanning tree $T$ of the graph $G$ is a tree subgraph of $G$ which does not contain any closed cycle
and includes all of the vertices. If we remove the spanning tree $T$ from the graph $G$, 
it consists of the disconnected pieces of the edges. 
After fixing the gauge,
the path integral is taken over the remaining $U_a$'s where the edge label $a$ runs over the edges on $G-T$.
We call the the number of the edges on $G-T$ as the rank $r$ of the graph $G$.

To perform the path integral over $U_a$ on $G-T$, it is useful to define the path representation of the Ihara zeta function. 
To this end, 
we need to introduce some additional notations: 
As we have explained, the index $a=1,\cdots,r$ corresponds to the edges in $G-T$. 
Corresponding to the inverse edges of them, we introduce the index $\bar{a}=r+1,\cdots,2r$ 
and we express them collectively as $\bsa \equiv (a,\bar{a})$ which runs from $1$ to $2r$. 
From this definition, it is natural to set $U_{\bar{a}}\equiv U_a^\dagger$. 
Since $T$ is a tree graph, such a reduced path $P_{\bsa\bsb}$ that starts from $e_\bsa \in G-T$ and ends at $s(e_\bsb)$ is unique and contains only the edges in $T$ except for $e_\bsa$.
Using these notations, the path representation of the Ihara zeta function is given by \cite{stark1999multipath}
\begin{equation}
  \zeta_G(q;U) = \det\left(1 - P_U\right)\,,
  \label{eq:path Uzeta}
\end{equation}
where $P_U$ is a square matrix of size $2r$ whose elements are given by 
\begin{equation}
  \left(P_U\right)_{\bsa\bsb} = 
  \begin{cases}
    q^{|P_{\bsa\bsb}|} U_{\bsa} & (\bsb \ne \bar{\bsa})\\
    0 & (\bsb = \bar{\bsa})
  \end{cases}\,.
  \label{eq:path matrix}
\end{equation}
Comparing to the vertex representation \eqref{eq:vertex Ihara} and the edge representation \eqref{eq:edge Ihara}, 
where the zeta function is described by the determinant of sparse matrices, 
the path representation is efficient for numerical simulations because the matrix $P$ is dense.

\section{The Functional Equation of the Ihara Zeta Function and the Riemann Hypothesis}
\label{sec:FuncEq}

In this section, we show that there is a dual expression of the Ihara zeta function of any connected graph in general, 
followed by recalling the functional equation of the Ihara zeta function of a regular graph. 
We will use the results in this section to show a duality in the FKM model in the following section. 

\subsection{Functional equation of the Ihara zeta function of a regular graph}
\label{subsec:regular graph}

Before discussing the functional equation of the Ihara zeta function, 
it is instructive to recall the more familiar Riemann zeta function, 
since the Ihara zeta function has many analogies to it.
The Riemann zeta function is defined by 
\be
\zeta(s) = \sum_{n=1}^\infty \frac{1}{n^s}\ ,
\ee
and it is shown that $\zeta(s)$ satisfies the following functional equation, 
\be
\zeta(s) = 2^s\pi^{s-1}\sin\left(\frac{\pi s}{2}\right)
\Gamma(1-s)\zeta(1-s)\ .
\ee
If we define the so-called the completed zeta function ($\xi$-function) as 
\be
\xi(s) \equiv \frac{1}{2}\pi^{-\frac{s}{2}}s(s-1)\Gamma\left(
\frac{s}{2}
\right)\zeta(s)\,,
\ee
it satisfies the symmetric functional equation, 
\be
\xi(s)=\xi(1-s)\, .
\label{symmetric functional equation}
\ee
This reflective symmetry is a key to understand 
that non-trivial zeros lies at least in the open strip of $0< \Real s < 1$,
which is called the {\it critical strip}. 
Furthermore, the conjecture that there exists
non-trivial zeros only on the line along $\Real s = \frac{1}{2}$ (critical line) is called the Riemann hypothesis.

The Ihara zeta function of a regular graph, where every vertex has the same degree, 
enjoys a similar relation to the functional equation of the Riemann zeta function. 
If we set the common degree by $\deg v=t+1$, 
the Ihara zeta function \eqref{eq:vertex Ihara} reduces to
\be
\zeta_G(q) = \left(1-q^2\right)^{-n_E+n_V}
  \det\left((1+tq^2)\bsone_{n_V}-q A \right)^{-1}\ .
  \label{eq:vertex Ihara regular}
\ee
Similar to the Riemann zeta function, we can show that the completed Ihara zeta function, 
\begin{align}
\xi_G(q)
&\equiv \left(1-q^2\right)^{n_E-\frac{n_V}{2}}
\left(1-t^2q^2\right)^{\frac{n_V}{2}}\zeta_G(q)\nn \\
&= 
\left(1-q^2\right)^{\frac{n_V}{2}}
\left(1-t^2q^2\right)^{\frac{n_V}{2}}
\det\left((1+tq^2)\bsone_{n_V}-q A \right)^{-1}\,,
\label{eq:completed zeta}
\end{align}
satisfies the symmetric functional equation,  
\be
\xi_G(q)
=(-1)^{n_V}\xi_G\left({1}/{t q}\right)\,.
\label{graph functional equation}
\ee
The proof is straightforward from the expression \eqref{eq:completed zeta}:
\begin{align}
  \xi_G(1/tq) 
  &= \left(1-\left(tq\right)^{-2}\right)^{\frac{n_V}{2}}
  \left(1-{q^{-2}}\right)^{\frac{n_V}{2}}
  \det\left(\left(1+(tq^2)^{-1}\right)\bsone_{n_V}-(tq)^{-1} A \right)^{-1} \nn \\
  &= \frac{(t^2q^2-1)^{\frac{n_V}{2}}(q^2-1)^{\frac{n_V}{2}}}{(tq^2)^{n_V}}
  \times(tq^2)^{n_V}
  \det\left(
  (tq^2+1)\bsone_{n_V} - qA
  \right)^{-1} \nn \\
  &= (-1)^{n_V}\xi_G\left({q}\right)\,.
\end{align}
If we change the variable $q$ with
\begin{equation}
q=t^{-s} \,,
\label{eq:q vs s}
\end{equation}
the functional equation (\ref{graph functional equation}) reduces to
\be
\xi_G(s)
=(-1)^{n_V}\xi_G\left(1-s\right). 
\ee
This is nothing but an analog of the symmetric functional equation (\ref{symmetric functional equation}) of the Riemann zeta
function.

For the $(t+1)$-regular graph, 
the radius of the largest circle of convergence $R_G$ with respect to $q$ is given by $1/t$ 
and all the poles exist in the region $1/t \le |q| \le 1$, that is, $0\le s\le 1$ 
which is consistent with the functional relation 
(For a proof, see Appendix~\ref{app:poles})%
\footnote{
In the Ihara zeta function,
we ara interested in the non-trivial poles instead of the non-trivial zeros.
}. 
It is also interesting that the Ihara zeta function $\zeta_G(q)$ of the $(t+1)$-regular graph 
has the non-trivial poles only on the critical line ($\Real s=1/2$), namely
it satisfies the Riemann hypothesis,
iff $G$ is Ramanujan, that is, 
the maximal eigenvalue $\lambda_{\rm max}$ except for $\lambda=t+1$ of the adjacency matrix $A$ satisfies%
\footnote{
Since the adjacency matrix $A$ satisfies $\sum_{v'=1}^{n_V}A_{vv'}=\deg v$, 
$A$ for a $(t+1)$-regular graph has a trivial eigenvector $(1,\cdots,1)^T$ with the eigenvalue $t+1$. 
This means that the Ihara zeta function of the $(t+1)$-regular graph has always poles at $q=1$ and $q=1/t$. 
} 
\begin{equation}
  |\lambda_{\rm max}| \le 2\sqrt{t}\,. 
  \label{eq:Ramanujan}
\end{equation}
In contrast to that the original Riemann hypothesis of the Riemann zeta function is highly non-trivial, 
we can easily prove it as follows: 

{\it 
Since the Ihara zeta function of the $(t+1)$-regular graph is expressed as \eqref{eq:vertex Ihara regular}, the non-trivial poles of $\zeta_G(q)$ is written down by using the eigenvalue $\lambda$ of the adjacency matrix $A$ as 
$q_\pm \equiv t^{-s_\pm} = \frac{\lambda \pm \sqrt{\lambda^2-4t}}{2t}$, 
which obey $t^{-s_+}t^{-s_-}=t^{-1}$ in general.  
We assume that $\lambda$ is not $t+1$ which is a trivial eigenvalue of $A$. 
When $\lambda^2 - 4t < 0$, $s_\pm$ become complex conjugate with each other and
these solutions satisfy $t^{-s_+}t^{-s_-}=t^{-2\Real s_\pm}$,
that means
the non-trivial poles lie along the line of $\Real s_\pm = 1/2$
on $s$-plane.  \hfill $\Box$
}

\begin{figure}[ht]
\begin{center}
\subcaptionbox{Tetrahedron $K_4$}[.45\textwidth]{
\includegraphics[scale=0.7]{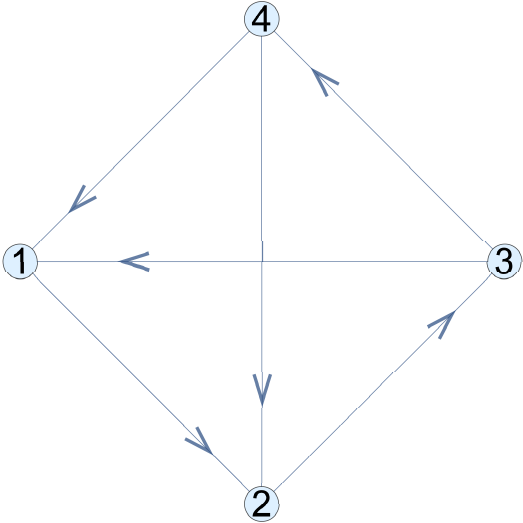}
}
\subcaptionbox{Double triangle $K_4-e$}[.45\textwidth]{
\includegraphics[scale=0.7]{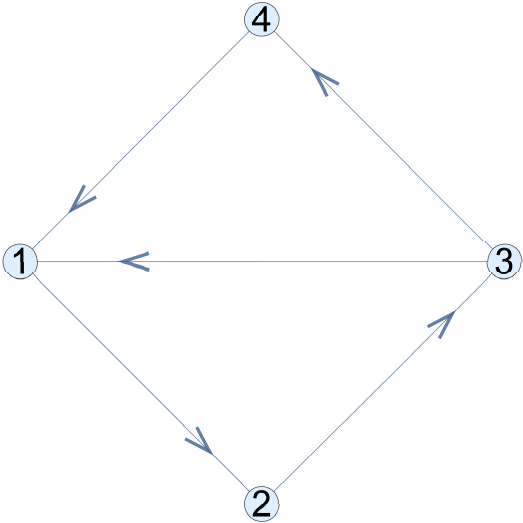}
}
\end{center}
\caption{Two graphs with four vertices. The double triangle
$K_4-e$ (DT) is obtained by removing one edge
from the tetrahedron $K_4$ of the complete graph.
$K_4$ is the regular graph with the same degree 3 for each vertex,
but DT is not regular.}
\label{K4 and DT graph}
\end{figure}

\begin{figure}[ht]
\begin{center}
\subcaptionbox{Tetrahedron $K_4$}[.45\textwidth]{
\includegraphics[scale=0.23]{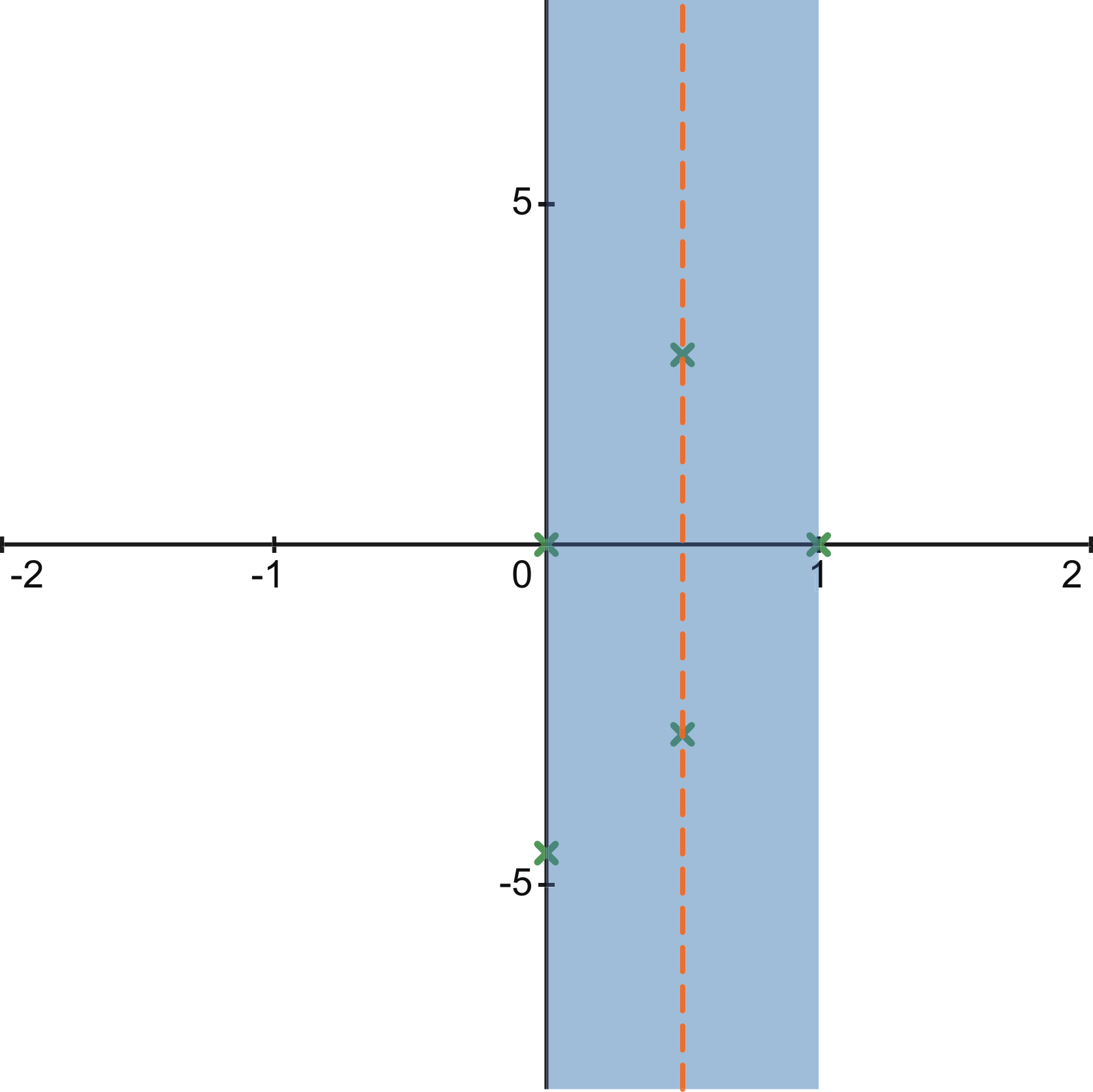}
}
\subcaptionbox{Double triangle $K_4-e$}[.45\textwidth]{
\includegraphics[scale=0.23]{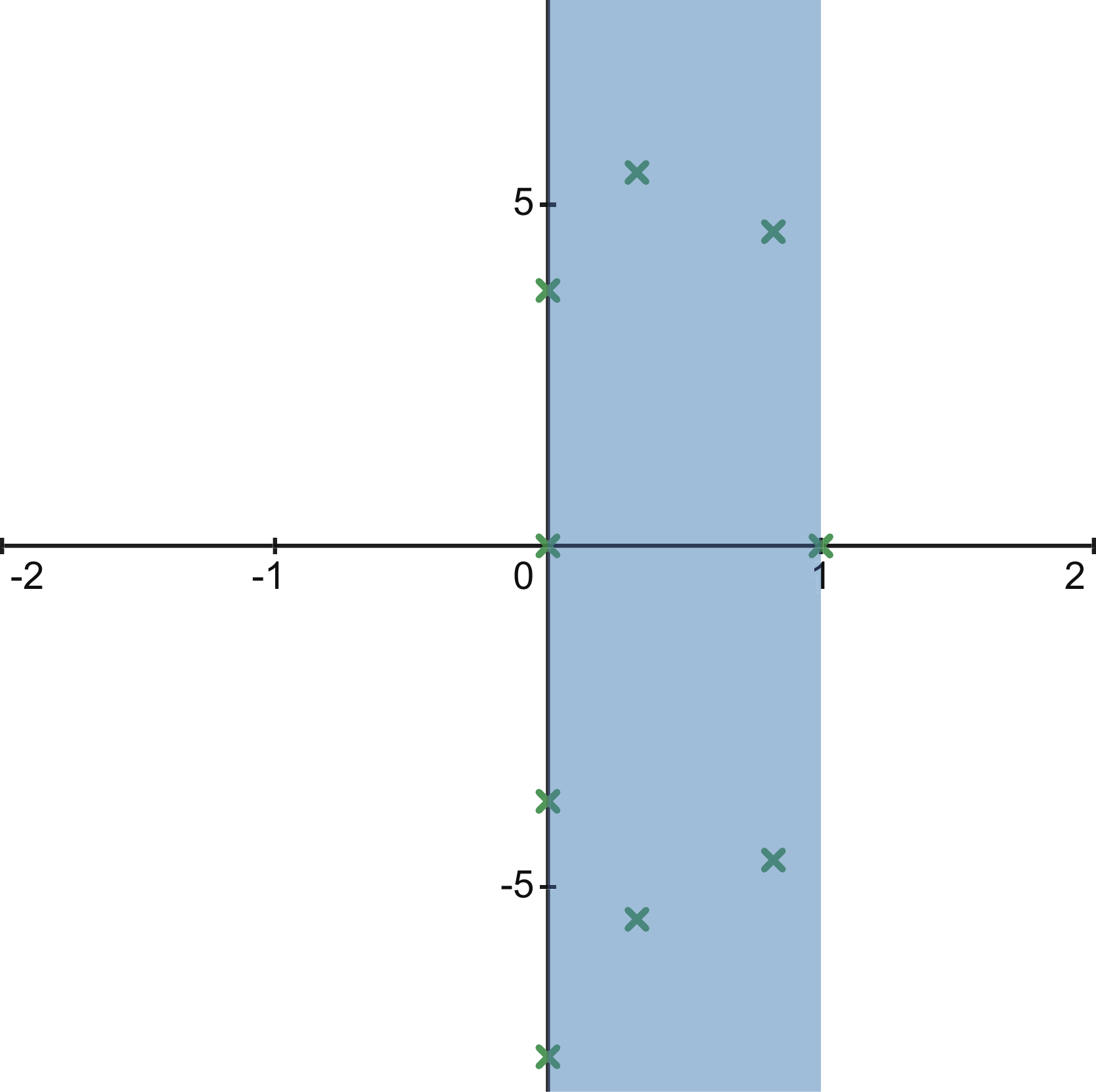}
}
\end{center}
\caption{The poles of the Ihara zeta function of the $K_4$ 
and $K_4-e$ (DT) graph on the $s$-plane.
The $K_4$ case satisfies the Riemann hypothesis while $K_4-e$ does not,
but all the poles exist in the critical strip with the boundary $0 \leq \Real s \leq 1$.}
\label{poles on s-plane}
\end{figure}

As a concrete example of a regular graph, 
let us consider the complete graph on four vertices $K_4$, that is, 
the tetrahedron graph
(see Fig.~\ref{K4 and DT graph} (a)).
The Ihara zeta function is given by
\be
\zeta_{K_4}(q) =\frac{1}{(1-q) (1-2 q)(1-q^2)^2 (1+q+2 q^2)^3}\,,
\label{K4 Ihara}
\ee
and the convergence radius is $R_{K_4}=1/2$ since $t=2$ in this case.
We see that the eigenvalues of the adjacency matrix are $(3,-1,-1,-1)$, 
and the maximal eigenvalue except for $t+1=3$ is $\lambda_{\rm max}=-1$ 
which satisfies $|\lambda|\leq 2\sqrt{2}$.
This means that the $K_4$ graph is Ramanujan.
The poles of the zeta function (\ref{K4 Ihara}) exist at
$q=\pm 1, -\frac{1\pm i \sqrt{7}}{4}, \frac{1}{2}$. Putting these poles on the $s$-plane as in  Fig.~\ref{poles on s-plane} (a), we see that the non-trivial poles ($q=-\frac{1\pm i \sqrt{7}}{4}$ and $\frac{1}{2}$)
lie along the critical line $\Real s=1/2$, that is,
the Riemann hypothesis is satisfied.

\subsection{Relations in the Ihara zeta function of an irregular graph}
\label{subsec:irregular graph}

We next consider the Ihara zeta function of an irregular graph $G$. 
In this case, the value of the convergence radius $R_G$ must be calculated explicitly for each graph, 
but the lower and the upper bound of $R_G$ is known as
\cite{Kotani2000ZetaFO}
\be
\frac{1}{t_{\text{max}}} \leq R_G \leq \frac{1}{t_{\text{min}}},
\label{eq:range RG}
\ee
where
$t_{\text{min}} = \min_{v\in V}(\deg v-1)$ and
$t_{\text{max}} = \max_{v \in V}(\deg v-1)$
(See Appendix~\ref{app:poles} for a proof).
Note that, if $G$ is the $(t+1)$-regular graph,
$R_G = {1}/{t}$ is reproduced from this relation 
because $t_{\rm max}=t_{\rm min}=t$. 
Since we can show that all the poles of $\zeta_G(q)$ exist in the range $R_G\leq |q| \leq 1$ also for an irregular graph (see Appendix \ref{app:poles}), it is reasonable to define a parameter $s$ as 
\begin{equation}
  q \equiv R_{G}^s\,, 
  \label{eq:q vs s general}
\end{equation}
corresponding to \eqref{eq:q vs s} for the regular graph.
Then, all the poles again exist in the critical strip with the boundary $0\leq \Real s \leq 1$ in this parametrization as well as for regular graphs. 

The most important difference from the regular graph is that there is no functional relation for the graph zeta function of the irregular graph.
Nevertheless, we can show that the Ihara zeta function with $q$ replaced by the inverse is again essentially the reciprocal of a polynomial of $q$ up to a non-essential factor. 
This can be seen from the fact that the Ihara zeta function satisfies the following identity in general
\begin{align}
  \zeta_G\left(R_G/q\right)^{-1} 
  &=
  \frac{\left( (q/R_G)^2 - 1 \right)^{n_E-n_V}\prod_{v\in V}(\deg v-1)}{(q/R_G)^{2n_E}}
   \nn \\
  &\hspace{2cm}\times 
  \det \left(
  {\bf 1}_{n_V}-q (R_G Q)^{-1}A+R_G^{-1} q^2(R_G Q)^{-1}
  \right)\,.
  \label{eq:dual trans of zeta}
\end{align}
The proof is straightforward as follows:

  {\it 
By changing the argument of the Ihara zeta function \eqref{eq:vertex Ihara} $q\to 1/cq$ with a parameter $c\ne 0$, we obtain
\begin{align}   
  \zeta_G(1/cq)^{-1} 
  &= \bigl(1-(cq)^{-2}\bigr)^{n_E-n_V}
  \det\bigl(\bsone_{n_V}-(cq)^{-1} A + (cq)^{-2} Q \bigr) \nn \\
  &= \bigl(1-(cq)^{-2}\bigr)^{n_E-n_V}
  \left(\det((cq)^{-2}Q)\right)
  \det\bigl(\bsone_{n_V}-cq Q^{-1} A + (cq)^{2} Q^{-1})\bigr) \nn \\
  &= 
  \frac{\left( c^2q^2 - 1 \right)^{n_E-n_V}\prod_{v\in V}(\deg v-1)}{(cq)^{2n_E N_c}}
  \det \left(
  {\bf 1}_{n_V N_c}-c q Q^{-1}A+c^2 q^2Q^{-1}
  \right)\,,
\end{align}
which reduces to \eqref{eq:dual trans of zeta} by setting $c=1/R_G$ as a special case of the identity. 
Note that $Q$ is invertible since we assume that the degrees are more than or equal to two. \hfill $\square$ }

\noindent
Note that,
if the graph is $(t+1)$-regular, the identity \eqref{eq:dual trans of zeta} becomes
\begin{align}
  \zeta_G\left(1/tq \right)^{-1} 
  &= 
\frac{(1-t^{-2}q^{-2})^{n_E-\frac{n_V}{2}}(1-q^{-2})^{\frac{n_V}{2}}}{(1-q^2)^{n_E-\frac{n_V}{2}}(1-t^2q^2)^{\frac{n_V}{2}}}
\zeta_G(q)^{-1} \,,
\label{eq:original duality}
\end{align}
which is equivalent to the functional relation \eqref{graph functional equation}. 

The structure of the right-hand side of \eqref{eq:dual trans of zeta} is essentially the same as the vertex representation of the Ihara zeta function \eqref{eq:vertex U-Ihara}. 
In this sense, \eqref{eq:dual trans of zeta} is a generalization of the functional relation of the Ihara zeta function. 
This transformation makes us possible to expand the graph zeta function by $1/q$
and we will use it to show an approximate duality of the FKM model on irregular graphs in the next section.

As an example of the irregular graph, let us see the double triangle (DT) which is obtained by removing one edge
from the tetrahedron $K_4$ (Fig.~\ref{K4 and DT graph} (b)).
For DT, the Ihara zeta function becomes
\be
\zeta_{\rm DT}(q) = 
\frac{1}{ \left(1-q^4\right) \left(1+q^2-2q^3\right)\left(1-q^2-2q^3\right)}\ .
\label{DT Ihara}
\ee
From this expression, we find $R_{\rm DT}=0.65729\cdots$, which satisfies
$1/t_{\rm max}<R_{\rm DT}<1/t_{\rm min}$ with $t_{\text{min}}=1$ and $t_{\text{max}}=2$.
We plot the poles in the $s$-plane in Fig.~\ref{poles on s-plane} (b). 
We see that all the poles exist in the critical strip with the boundary $0\leq \Real s \leq 1$ as expected,
but they clearly do not satisfy the Riemann hypothesis.

\section{Duality of the FKM Model and the GWW Phase Transitions}
\label{sec:duality} 

\subsection{Stability of the FKM model}
\label{subsec:Stability}

We first would like to discuss the stability of the model on a general graph.  

Since the action of the FKM model \eqref{eq:action} is Gaussian with respect to $\Phi_v^I$,
the matrix
$\Delta(q;U) \equiv {\bf 1}_{N_c n_V} - q A_U +q^2Q$ must be positive definite in order for the system to be stable. 
Since we are interested in the stability around the vacuum, namely, $U_a={\bf 1}_{N_c}$ ($a=1,\cdots,r$) after fixing the gauge, 
we need only to examine the positive definiteness of the matrix of size $n_V$, 
\begin{equation}
  \Delta(q) \equiv \Delta(q;U=1) = {\bf 1}_{n_V} - q A +q^2 Q\,,
\end{equation}
which is called the deformed graph Laplacian. 

Let us consider the behavior of the eigenvalues $\lambda_i(q)$ ($i=1,\cdots,n_V$) of $\Delta(q)$ for $q>0$. 
The point is that the product of the eigenvalues is essentially
expressed by the inverse of the Ihara zeta function, 
\begin{equation}
  \det\Delta(q) = \prod_{i=1}^{n_V}\lambda_i  = (1-q^2)^{-n_E+n_V} \zeta_G(q)^{-1}\,.
\end{equation}
Let us first consider the region $0<q<R_G$.
Since $\Delta(q) \to {\bf 1}_{n_V}$ in the limit of $q\to 0$, all the eigenvalues $\{\lambda_i(q)\}$ ($i=1,\cdots,n_V$) satisfy $\lim_{q\to 0}\lambda_i(q)= 1>0$.
According to 
Kotani-Sunada's theorem \cite{Kotani2000ZetaFO},
$\zeta_G(q)$ has a first pole at $q=R_G<1$ which has the smallest magnitude of all poles.
This means that the sign of one of the eigenvalues of $\Delta(q)$ (we call it $\lambda_1(q)$) 
flips at $q=R_G$.
Therefore, the vacuum $U_a=\bsone_{N_c}$ is stable in $0<q<R_G$ while it becomes unstable at $q= R_G$.

On the other hand,
since $\lim_{q\to\infty}\Delta(q) = q^2 Q$, 
the eigenvalues $\lambda_i(q)$ $(i=1,\cdots,n_V)$ asymptotically behave as $\lim_{q\to\infty}\lambda_{i}(q) = q^2 (\deg v-1)$ for ${}^{\exists}v\in V$. 
Therefore, all the eigenvalues $\lambda_i(q)$ are positive for $q>1$ because $\zeta_G(q)$ has no pole in $q>0$, that is, the vacuum $U_a=\bsone_{N_c}$ is stable in $q>1$. 

The matrix $\Delta(q)$ at $q=1$,   
\begin{equation}
  \Delta(1)= \bsone_{n_V} -A + Q \,,
\end{equation}
is nothing but the graph Laplacian. 
As discussed in \cite{Ohta:2021tmk}, the graph Laplacian of a connected graph has only one zero mode, $\left(1,\cdots,1\right)^T$, since it is expressed by the incidence matrix%
\footnote{
This expression of the incidence matrix is possible since we assume that the graph is simple. 
}, 
\begin{equation}
L^e_{\ v} = 
\begin{cases}
  1 & v=t(e) \\
  -1 & v=s(e) \\
  0 & {\rm others}
\end{cases}\,,
\label{eq:incidence matrix}
\end{equation}
as
\begin{equation}
 \Delta  = L^T L \,,
\end{equation}
and $\dim\ker L=1$ for a connected graph
(See also \cite{yumoto2022lattice,yumoto2023new, Yumoto:2023hnx}). 
Therefore, one of the eigenvalues becomes zero at $q=1$,
namely
the FKM model becomes unstable at $q=1$. 
Combining these results, we can conclude that the vacuum $U_a={\bf 1}_{N_c}$ is stable in the regions $0<q<R_G$ and $q>1$, 
but it is unstable at least in the vicinity of the boundaries of $R_G<q<1$. 

Whether there are other stable regions in $R_G<q<1$ or not depends on the positions of the real poles of $\zeta_G(q)$. 
If $\zeta_G(q)$ has no real pole in $0<q<1$ except for $q=R_G$, 
the corresponding FKM model is unstable in the whole region of $R_G<q<1$, namely in the critical strip, because there is no chance to flip the sign of the eigenvalue $\lambda_1(q)$ other than $q=R_G$ and $q=1$. 
Since all of the graphs we consider in the next section have this property, the FKM models we numerically simulate in this paper are unstable in the critical strip.
However, this is not always the case, as real poles can generally exist in the region $R_G<q<1$.
For graphs that have real poles in the range $0<q<1$ other than $q=R_G$, there can be extra stable regions in the middle. 
Of course, even in that case, it remains true that all unstable regions exist in the critical strip.

We note that, in the numerical simulations shown in Secs.~\ref{sec:Regular} and \ref{sec:Irregular}, the system shows instability when the value of $q$ is not in the critical strip but close to its boundaries, that is, $q\lesssim R_G$ and $q\gtrsim 1$ ($s\gtrsim 1$ and $s\lesssim 0$, equivalently). 
This is due to the fluctuation of the unitary matrices. 
In fact, the above discussion of the stability is based on the assumption that all the unitary matrices are set to the unit matrix. 
Although the stability of the model with non-trivial configurations of the unitary matrices remains an unresolved issue, it is reasonable to assume that these matrices can easily transition to an unstable region due to the tunneling effect, particularly when the coupling constant approaches to the critical strip.
This is thought to be the reason why the numerical simulations around the critical strip fail.

\subsection{Duality of the FKM Model}
\label{subsec:duality} 
Since the partition function of the FKM model is described by the unitary matrix weighted Ihara zeta function,
the functional equation of the Ihara zeta function
is carried over to the FKM model as a duality between
the small and large couplings. 
In fact, 
the proof of the transformation \eqref{eq:dual trans of zeta} is also applicable to the unitary weighted Ihara zeta function and the following identity for the matrix weighted Ihara zeta function holds%
\footnote{
$Q$ has been again redefined to $Q=\diag(t_1\bsone_{N_c},t_2\bsone_{N_c},
\cdots, t_{n_V}\bsone_{N_c})$.
}
\begin{align}
\zeta_G\left({R_G}/{q};U\right)^{-1}
&=
  \frac{\left( (q/R_G)^2 - 1 \right)^{(n_E-n_V)N_c}\prod_{v\in V}(\deg v-1)^{N_c}}{(q/R_G)^{2n_E N_c}}
   \nn \\
  &\hspace{2cm}\times 
  \det \left(
  {\bf 1}_{n_V}-q (R_G Q)^{-1}A_U+R_G^{-1} q^2(R_G Q)^{-1}
  \right)\,. 
\label{zeta function transf}
\end{align}

Similar to the usual Ihara zeta function, 
the relation (\ref{zeta function transf}) reduces to the functional equation of the matrix weighted Ihara zeta function
\begin{align}
  \zeta_G\left(1/tq;U \right)^{-1} 
  &= 
\frac{(1-t^{-2}q^{-2})^{n_E-\frac{n_V}{2}}(1-q^{-2})^{\frac{n_V}{2}}}{(1-q^2)^{n_E-\frac{n_V}{2}}(1-t^2q^2)^{\frac{n_V}{2}}}
\zeta_G(q;U)^{-1} \,,
\label{vertex duality}
\end{align}
when $G$ is $(t+1)$-regular.
Since the effective action of the FKM model is also written in terms of the unitary matrix weighted Ihara zeta function as \eqref{eq:Seff}, 
there exists the duality on the couplings between $q$ and $1/t q$ for the FKM model on the regular graph,
\begin{equation}
  S_{\rm eff}(1/tq;U) = S_{\rm eff}(q;U)\,,
  \label{eq:exact duality}
\end{equation}
up to an irrelevant constant shift. 
Therefore, when $G$ is regular, the dynamics of the FKM model in the region $q>1$ is identical with that in $0<q<R_G$ even if the series expansion of the effective action \eqref{eq:Seff} converges only in $0<q<R_G$.

On the other hand, when $G$ is irregular, the right-hand side of the transformation \eqref{vertex duality} cannot be interpreted as a unitary matrix weighted Ihara zeta function.
Therefore, the dynamics of the FKM model in $q>1$ is not the same with that in the region $0<q<R_G$. 
Nevertheless, the effective action of the FKM model in $q>1$ can be expressed in a sum of all possible Wilson loops on $G$. 
In fact, changing the argument of \eqref{zeta function transf} from $q$ to $R_G/q$, we obtain
\begin{multline}
  \zeta_G(q;U)^{-1}
  = \frac{\left( q^{-2} - 1 \right)^{(n_E-n_V)N_c}\prod_{v\in V}(\deg v-1)^{N_c}}{q^{-2n_E N_c}}\\
  \times\det \left(
  {\bf 1}_{n_V}-q^{-1}  Q^{-1}A_U+ q^{-2} Q^{-1}
  \right)\,. 
\end{multline}
This means that the effective action \eqref{eq:Seff} of the FKM model can also be written as 
\begin{equation}
  S_{\rm eff}(q;U) = -\gamma N_c \log\zeta_G(q;U)
  = \gamma N_c \Tr\log\left(
  {\bf 1}_{n_V}-q^{-1}  Q^{-1}A_U+ q^{-2} Q^{-1}
  \right)\,,
  \label{eq:Seff dual}
\end{equation}
up to irrelevant constant terms in general. 
In particular, since the series expansion by $q^{-1}$ of the right-hand side becomes well-defined for large $q$,
we can interpret again the effective action of the FKM model in a sum of all possible Wilson loops on $G$ as announced.
Of course, the detail of the expansion depends on $Q$ which gives non-trivial weights to the expansion when $G$ is irregular. 
Therefore, to see the difference of the behavior of the FKM model in the region $q>1$ from that in the region $0<q<R_G^{-1}$, we need to carry out numerical simulations. 
We will see it in the next section in concrete examples.

\subsection{Phase structure of the FKM model on the cycle graph}
\label{subsec:cycle graph}

Before considering the phase structure of the model on a general graph, 
it is instructive to recall the exact results of the FKM model on the cycle graph whose effective action is given by 
\begin{equation}
  S_{\rm eff}^{C_n}(q;U)=-\gamma N_c \Tr\left[\sum_{m=1}^\infty \frac{q^{nm}}{m}\left( U^m +  U^{-m}\right)\right]\,.
  \label{eq:effective action Cn}
\end{equation}
In \cite{PhysRevD.108.054504}, the partition function (the free energy) of the model in $0<q<1$ of this system is analytically evaluated. 
In this subsection, we will explain the complete phase structure of the cycle graph by combining this exact analysis with the duality discussed in the previous subsection. 

\begin{figure}[H]
  \begin{center}
  \includegraphics[scale=0.6]{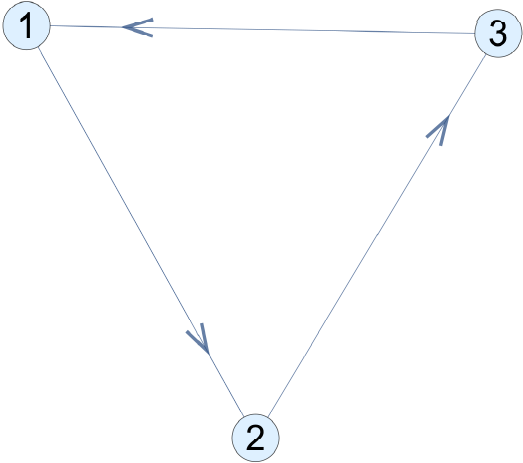}
  \end{center}
  \caption{A cycle graph with three vertices.}
  \label{C3 graph}
\end{figure}
The cycle graph $C_n$ is a 2-regular graph with $n$ vertices $\{v_i\}$ and $n$ edges; $e_i=\langle v_i, v_{i+1}\rangle$
for $i=1,\cdots,n-1$ and $e_n=\langle v_n, v_1\rangle$. (See e.g. $C_3$ in Fig.~\ref{C3 graph}).
Since a maximal tree of $C_n$ can be set to $T=e_2\cdots e_n$, 
we can fix the gauge as $U_{e_2}=\cdots=U_{e_n}=\bs{1}_{N_c}$. 
Therefore, the rank of $C_n$ is $1$ and only the unitary matrix on $e_1$, $U \equiv U_{e_1}$, survives after the gauge fixing. 
Then, the partition function is expressed in terms of the eigenvalues $e^{i\theta}$ ($i=1,\cdots,N_c$) of the unitary matrices $U$ as 
\begin{align}
  Z_{C_n} 
  &= {\cal N} \int_{-\pi}^{\pi} \prod_{i=1}^{N_c} d\theta_i\,
  e^{\sum_{j\ne k}\log\left|\sin \frac{\theta_j-\theta_k}{2}\right|
   -N_f\sum_{i} \log\left(1-2q^n \cos\theta_i + q^{2n} \right)
  }\,,
  \label{eq:ZCn}
\end{align}
where the first term $\sum_{j\ne k}\log\left|\sin \frac{\theta_j-\theta_k}{2}\right|$ arises from the Jacobian associated with diagonalization of $U$ and ${\cal N}$ is an irrelevant constant. 
We note that this expression is valid only in $0<q<1$ because the expansion of $U$ in the effective action is defined only in this region. 

\begin{figure}[h]
  \begin{center}
  \subcaptionbox{$\alpha=0.01$}[.45\textwidth]{
  \includegraphics[scale=0.45]{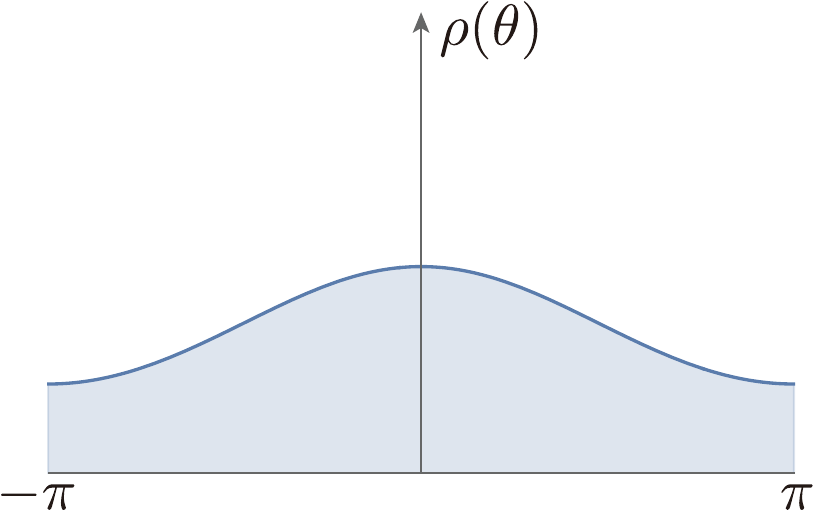}
  }
  \hspace*{0.5cm}
  \subcaptionbox{$\alpha=0.08$}[.45\textwidth]{
  \includegraphics[scale=0.45]{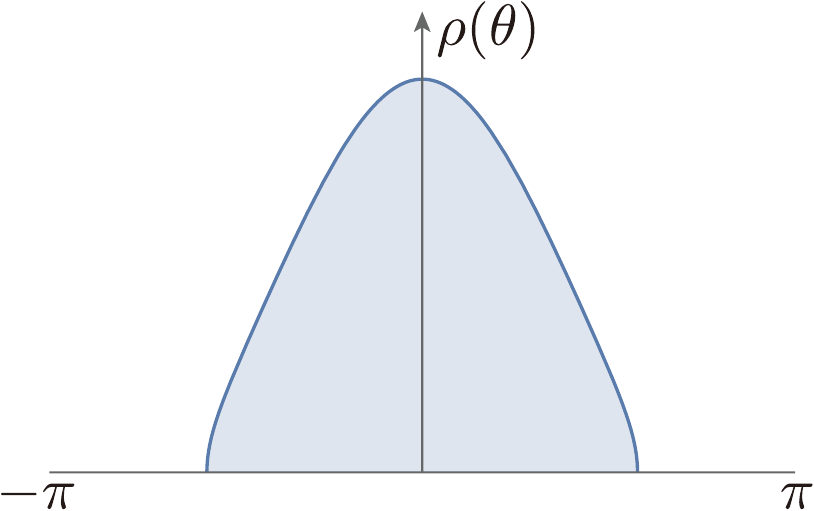}
  }
  \end{center}
  \caption{
  The eigenvalue distributions of the FKM model on $C_n$ with $\gamma=10$, that is, ${q^*}^n=\frac{1}{19}\simeq0.053$. 
  (a)
  For $q<{q^*}$, the eigenvalue distribution is continuous on the cycle. 
  (b)
  For ${q^*}<q$, the eigenvalue distribution is concentrated around $\theta=0$.} 
  \label{Distribution of Eigenvalue}
\end{figure}

In the limit of $N_c\to\infty$ with fixing $\gamma = N_f/N_c$, 
the saddle point approximation becomes exact and
the eigenvalues $\theta_i$ can be regarded as a continuous variable $\theta(x)$ $(x\in[0,1])$. 
As a result, the density of the eigenvalues, 
\begin{equation}
  \rho(\theta)\equiv \frac{1}{N_c}\sum_{i=1}^{N_c}\delta(\theta-\theta_i)\,,
\end{equation}
can be evaluated exactly as 
\begin{equation}
  \rho(\theta) = \begin{cases}\displaystyle
    \frac{1}{2\pi}\left(1+2\gamma\frac{q^n \cos\theta-q^{2n}}{1-2 q^n \cos\theta+q^{2n}}\right)\,, & \quad (\theta_0=\pi) \vspace{4mm}\\
    \displaystyle
    \frac{2(\gamma-1) q^n }{\pi}\frac{\cos\frac{\theta}{2}}{1-2 q^n \cos\theta+q^{2n}}\sqrt{\sin^2\frac{\theta_0}{2}-\sin^2\frac{\theta}{2}}\,, & \quad (\theta_0<\pi)
  \end{cases}
  \label{eq:sln}
\end{equation}
where $\theta_0$ is determined by 
\begin{equation}
  \sin^2\frac{\theta_0}{2} = \frac{(1-q^n)^2}{4q^n}\frac{2\gamma-1}{(\gamma-1)^2}\,,
\end{equation}
which expresses the boundary of the support of the density function, namely, $\rho(\theta)\ge 0$ in the region $-\theta_0\le \theta \le \theta_0$ and $\rho(\theta)=0$ outside. 
In other words, 
$\theta_0 = \pi$ means the eigenvalue distribution is continuous on the cycle $e^{i\theta}$, 
whereas $\theta_0<\pi$ means that the eigenvalues are distributed over only a portion of the circle around $\theta=0$
(See Fig.~\ref{Distribution of Eigenvalue}). 
Since $\theta_0=\pi$ is equivalent to
\begin{equation}
  (q^*)^n = \frac{1}{2\gamma-1}\,,
  \label{eq:q critical}
\end{equation}
which gives the critical coupling $q = {q^*}$.  
This transition indexed by the eigenvalue distribution is the so-called GWW phase transition \cite{Gross:1980he,Wadia:1980cp}. 
For more detail, see \cite{PhysRevD.108.054504}.

The free energy of the system is defined by
\begin{align}
  F_{G} &\equiv -\lim_{N_c\to\infty} \frac{1}{N_c^2}\log Z_{G}\,.
  \label{eq:free energy}
\end{align} 
In order to evaluate the order of the phase transition, 
it is also convenient to define the ``internal energy'', 
\begin{equation}
  E_G \equiv \gamma \frac{\partial F_{G}}{\partial \gamma}\,,
  \label{eq:energy}
\end{equation}
the ``specific heat'', 
\begin{equation}
  C_G \equiv -\gamma^2 \frac{\partial^2 F_{G}}{\partial \gamma^2}\,, 
  \label{eq:specific heat}
\end{equation}
and the ``derivative of the specific heat'', 
\begin{equation}
  dC_G \equiv \gamma^3 \frac{\partial^3 F_{G}}{\partial \gamma^3}\,, 
  \label{eq:dC}
\end{equation}
by regarding the parameter $\gamma$ as inverse temperature.

From the eigenvalue density \eqref{eq:sln}, the free energy of the FKM model on $C_n$ can be evaluated as \cite{PhysRevD.108.054504} 
\begin{equation}
F_{C_n} = \begin{cases}
\gamma^2 \log(1-q^{2n}) & (0< q \leq {q^*})\\
(2\gamma-1)\log(1-q^n)+\frac{1}{2}\log (2 \gamma q^n) +f_1(\gamma)
& ({q^*}< q < 1)
\end{cases}\, ,
\label{eq:free energy Cn}
\end{equation}
where
\begin{equation}
  f_1(\gamma) = 
  (\gamma-1)^2\log\left(1 - \frac{1}{\gamma}\right) 
  -2\left(\gamma - \frac{1}{2}\right)^2\log\left(1 - \frac{1}{2\gamma}\right) 
\end{equation}
is determined by the condition that the free energy is continuously connected at the critical coupling $q={q^*}$.
Substituting the expression \eqref{eq:free energy Cn} into the definitions \eqref{eq:energy}, \eqref{eq:specific heat}
and \eqref{eq:dC}, 
we obtain the internal energy, 
\begin{equation}
  E_{C_n} = \begin{cases}
    \displaystyle 
    2\gamma^2 \log(1-q^{2n}) &  (0< q \leq {q^*})\\
    \displaystyle 
    2\gamma \log(1-q^n) + e_1(\gamma)
    & ({q^*}< q < 1)
  \end{cases}\,,
  \label{eq:energy Cn}
\end{equation}
where 
\begin{equation}
  e_1(\gamma) \equiv 
  2\gamma \left(\gamma-1\right)\log\left(1 - \frac{1}{\gamma}\right)
  - 2\gamma(2\gamma - 1)\log\left(1 - \frac{1}{2\gamma}\right) \,,
\end{equation}
the specific heat, 
\begin{equation}
  C_{C_n} = \begin{cases}
    \displaystyle -2\gamma^2 \log(1-q^{2n}) &  (0 < q \leq {q^*})\\
    \displaystyle -2\gamma^2 \log \frac{4\gamma(\gamma-1)}{(2\gamma-1)^2} & ({q^*} < q < 1)
  \end{cases}\,,
  \label{eq:specific heat Cn}
\end{equation}
and the derivative of the specific heat,
\begin{equation}
  dC_{C_n} = \begin{cases}
    0 & (0< q \leq {q^*})\\
    \displaystyle \frac{2\gamma^3}{2\gamma^3 - 3\gamma^2 + \gamma}
    & ({q^*}< q < 1)
  \end{cases}\,, 
  \label{eq:dC Cn}
\end{equation} 
for $G=C_n$.
It is easy to check that the internal energy \eqref{eq:energy Cn} 
and the specific heat \eqref{eq:specific heat Cn}
connect continuously at $q={q^*}$ 
while the derivative of the specific heat does not. 
This means that the GWW phase transition of the FKM model on $C_n$ is third-order as the same as the original GWW model%
\footnote{
In \cite{PhysRevD.108.054504}, it has been shown that the phase transition is third-order in the sense that the first and second derivative of the free energy in terms of the {\it coupling constant $\mathit q$}. 
It is interesting that the transition is still third-order in terms of $\gamma$ even if $q$ and $\gamma$ are independent in the FKM model. 
}. 

Although we have considered only the region $0<q<1$ so far, 
we can extend it to $q>1$ by using the small/large coupling duality explained in Sec.~\ref{subsec:duality}.
Since the cycle graph is 2-regular graph, it obeys the functional equation, 
\begin{equation}
  \zeta_{Cn}(1/q;U) = q^{-2nN_c} \zeta_{C_n}(q;U)\,. 
\end{equation}
This means that the free energy in the region $q>1$ is given by 
\begin{equation}
  F_{C_n}(1/q) = F_{C_n}(q) - 2n\gamma \log q\,.
  \label{eq:duality for Cn}
\end{equation}
In order to display the duality in a symmetric manner, it is convenient to change the variable from $q$ to $s$ as \eqref{eq:q vs s general}. 
However, we cannot adopt $q=R_G^{s}$ in this case because the cycle graph has exceptionally $R_G=1$, in other words, the critical strip shrinks to the line associated with $|q|=1$. 
We thus instead define 
\begin{equation}
  s = -\log q\,, 
  \label{eq:s Cn}
\end{equation}
for $G=C_n$.
In this parametrization, the original region $0<q<1$ and the dual region $q>1$ correspond to $s>0$ and $s<0$, respectively. 
Corresponding to \eqref{eq:q critical}, the critical coupling in terms of the parameter $s$ is 
\begin{equation}
  s^* = \frac{1}{n} \log(2\gamma-1)\,. 
  \label{PT point for Cn}
\end{equation}
Therefore, we expect that the FKM model has two GWW phase transition points at $s=\pm s^*$.

\begin{figure}[H]
  \begin{center}
  \subcaptionbox{$E_{C_n}$}[.30\textwidth]{
  \includegraphics[scale=0.22]{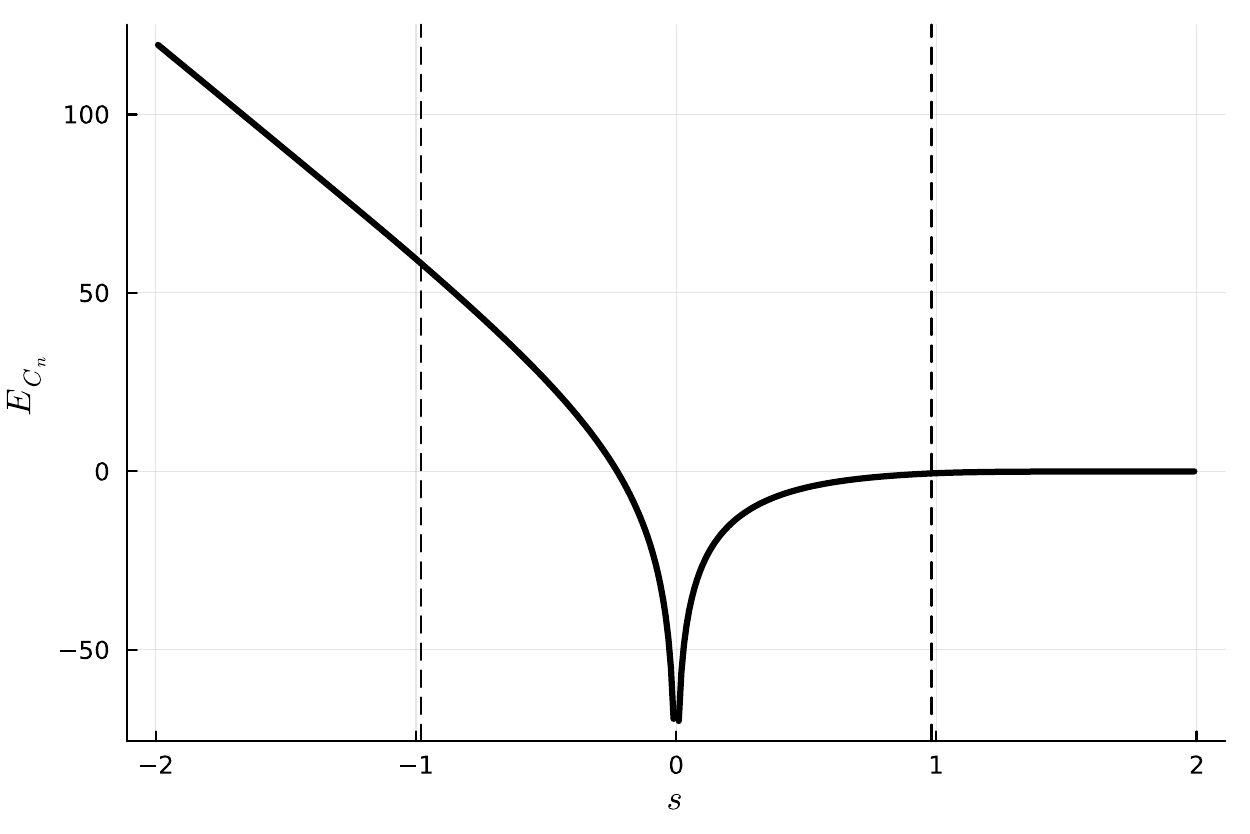}
  }
  \hspace*{0.5cm}
  \subcaptionbox{$C_{C_n}$}[.30\textwidth]{
  \includegraphics[scale=0.22]{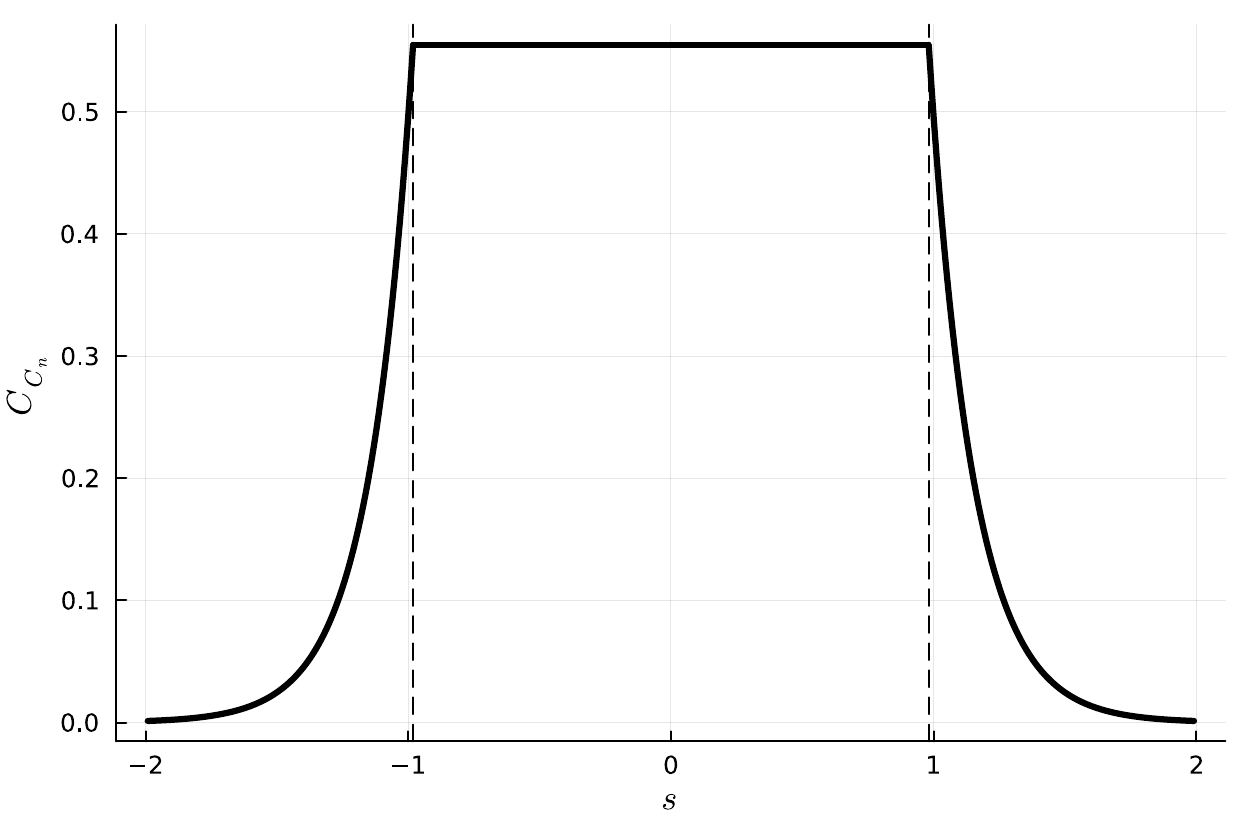}
  }
  \hspace*{0.5cm}
  \subcaptionbox{$dC_{C_n}$}[.30\textwidth]{
  \includegraphics[scale=0.22]{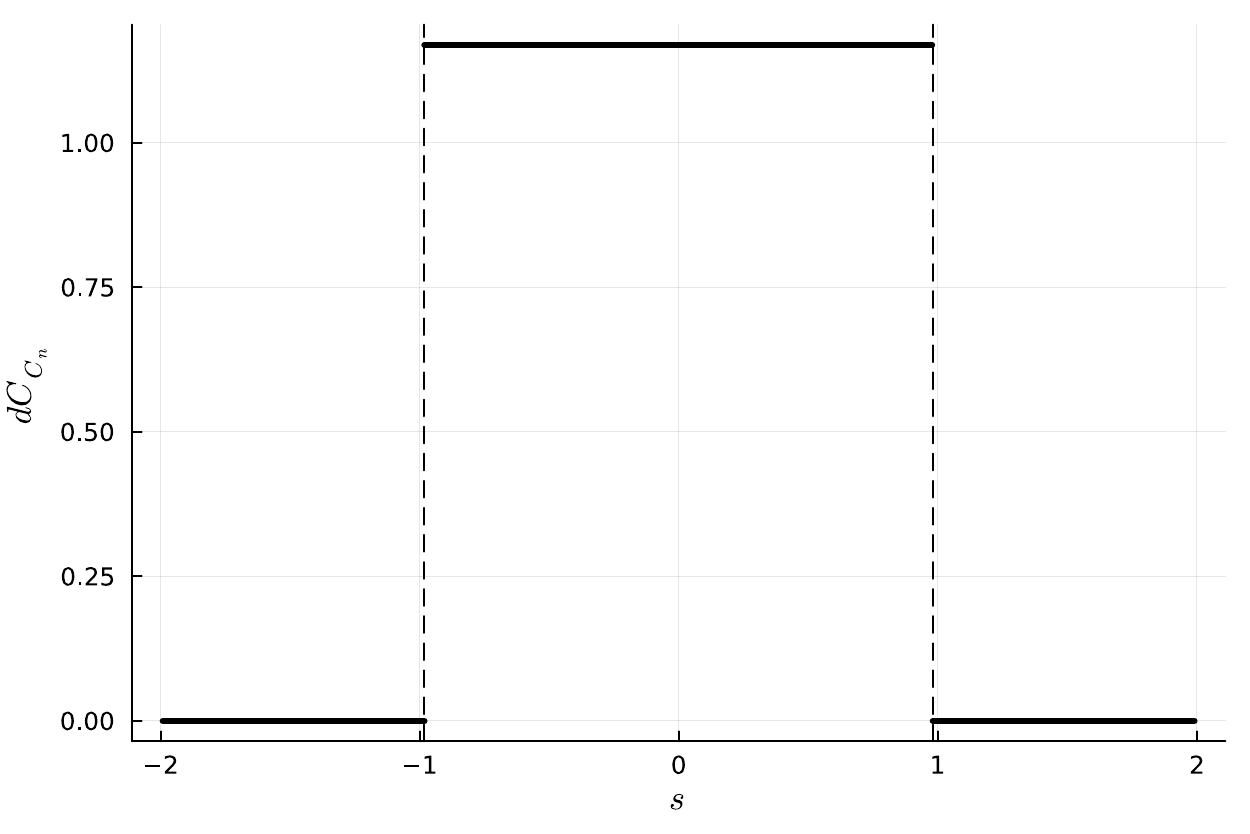}
  }
  \end{center}
  \caption{
  The behavior of the observables of the FKM model on $C_n$ with $n=3$, $\gamma=10$ and $N_c=16$. 
  The vertical dashed lines show the positions of the critical coupling $\pm s^* \simeq \pm 0.98$ in this parametrization. 
  (a) the internal energy, 
  (b) the specific heat, 
  (c) the derivative of the specific heat.
  }
  \label{fig:obs Cn}
\end{figure}

Taking into account the duality \eqref{eq:duality for Cn}, 
the free energy for all region of $s$ becomes
\begin{equation}
  F_{C_n} = \begin{cases}
  \gamma^2 \log(1-e^{-2ns}) & (s^* \le  s)\\
  (2\gamma-1)\log(1-e^{-ns})+\frac{1}{2}\log (2 \gamma e^{-ns}) +f_1(\gamma)
  & (0< s < s^*) \\
  (2\gamma-1)\log(1-e^{ns})+\frac{1}{2}\log (2 \gamma e^{ns}) + f_1(\gamma) - 2n\gamma s 
  & (-s^* < s < 0) \\
  \gamma^2 \log(1-e^{2ns}) -2n\gamma s & (s \le -s^*)
  \end{cases}\,.
  \label{eq:free energy Cn s}
 \end{equation}
 Similarly, the internal energy, the specific heat and the derivative of the specific heat become
 \begin{align}
  \label{eq:energy Cn s}
  E_{C_n} &= 
  \begin{cases}
    \displaystyle 
    2\gamma^2 \log(1-e^{-2ns}) &  (s^* \le  s)\\
    \displaystyle 
    2\gamma \log(1-e^{-ns}) + e(\gamma)
    & (0< s < s^*) \\
    \displaystyle 
    2\gamma \log(1-e^{ns}) + e(\gamma) -2n\gamma s
    & (-s^* < s < 0) \\
    \displaystyle 
    2\gamma^2 \log(1-e^{2ns}) - 2n\gamma s &  (s \le -s^*)
  \end{cases}\,, \\
  \label{eq:specific heat Cn s}
  C_{C_n} &= \begin{cases}
    \displaystyle -2\gamma^2 \log(1-e^{-2n|s|}) &  (s^* \le |s|)\\
    \displaystyle -2\gamma^2 \log \frac{4\gamma(\gamma-1)}{(2\gamma-1)^2} & (0< |s| < s^*)
  \end{cases}\,, \\
  \label{eq:dC Cn s}
  dC_{C_n} &= \begin{cases}
    0 & (s^* \le  |s|)\\
    \displaystyle \frac{2\gamma^3}{2\gamma^3 - 3\gamma^2 + \gamma}
    & (0< |s| < s^*)
  \end{cases}\,, 
\end{align}
respectively. 
We see that the order of the phase transition at $s=-s^*$ is also of the third-order as expected. 
We plot \eqref{eq:energy Cn s}, \eqref{eq:specific heat Cn s} and \eqref{eq:dC Cn s} in Fig.~\ref{fig:obs Cn} where we can see the order of the phase transitions and the duality clearly. 

\subsection{GWW phase transitions in the FKM model for large $\gamma$}
\label{subsec:GWW limit}

As shown in \cite{PhysRevD.108.054504}, 
the FKM model for $N_c\to\infty$ generally undergoes the GWW phase transitions in the region $0<q<R_G$ when $\gamma$ exceeds a value of ${\cal O}(1)$.
Since the GWW phase transition occurs when the distribution of eigenvalues of a Wilson loop (or a Polyakov loop) on $S^1$ breaks or connects in the process of changing the coupling constant, 
the number of the phase transitions in the FKM model on a general graph $G$ is essentially the same as the rank of $G$, that is, the number of fundamental cycles of $G$ which generate the Wilson loops.
However, as we will see shortly, the actual number of the phase transitions depends on the structure of the graph since some of the fundamental cycles simultaneously exhibit the GWW phase transition.

In order to analyze the phase structure of the FKM model on a general graph, it is instructive to look at the model on the cycle graph considered in the previous section in the limit, 
\begin{equation}
  \gamma\to\infty, \quad q\to 0, \quad 
  \gamma q^n \equiv \lambda^{-1} \,:\,{\rm fixed}\,. 
\end{equation}
In this limit, the effective action \eqref{eq:effective action Cn} reduces to 
\begin{equation}
  S_{\rm eff}^{C_n}(q;U)\to -\frac{N_c}{\lambda}\Tr\left(U + U^\dagger \right)\,,
\end{equation}
which is nothing but the action of the GWW model.
Not only the tree level action, 
but also the limit of the exact solution of the eigenvalue density \eqref{eq:sln} becomes that of the GWW model \cite{PhysRevD.108.054504}. 
Since the phase transition of the GWW model takes place at $\lambda=2$, the position of the phase transition of the FKM model for large $\gamma$ is estimated as 
\begin{equation}
  q^* \simeq \left(2\gamma\right)^{-1/n}\,,
\end{equation}
which is consistent with the exact result \eqref{eq:q critical}. 
In addition, the phase transition of the GWW model is third-order, which is the also consistent with the fact that the phase transition of the FKM model on $G=C_n$ is third-order. 
From this observation, we see that we can estimate the nature of the phase transition of the FKM model for large $\gamma$ when the effective action can be approximated by the GWW model. 

Whether the FKM model can be approximated by the GWW model or not depends on the structure of the graph. 
To see it, 
we adopt as the set of the fundamental cycles the one with the smallest length%
\footnote{
A choice of fundamental cycles is not unique in general. 
},
\begin{equation}
  {\cal F} \equiv \{[C_a]\,|\, a=1,\cdots,r,\ |C_1|\le\cdots\le|C_r|\}\,, 
  \label{eq:fundamental cycles}
\end{equation} 
and suppose that ${\cal F}$ consists of cycles of lengths $l_i$ ($i=1,\cdots,b_G$, $l_1<\cdots<l_{b_G}$). 
In other words, $b_G$ is the number of the types of polygons in the fundamental cycles of the least lengths.
Correspondingly, we define $b_G$ subsets of ${\cal F}$, 
\begin{equation}
  {\cal F}_i \equiv \{[C_a]\in {\cal F}\,|\, |C_a| = l_i\}, \quad m_i \equiv |{\cal F}_i|\,, 
  \label{eq:Fi}
\end{equation} 
which is the set of $l_i$-gons in $\cal{F}$. 
In addition, we define a subset of $[{\cal P}_R]$ whose length is $l_i$ but are not in ${\cal F}_i$ as 
\begin{equation}
  {\cal G}_i \equiv \left\{[C]\in [{\cal P}_R]\,| \,|C|=l_i,\ [C]\not\in {\cal F}_i  \right\}\,,
\end{equation}
whose elements are composite cycles by definition. 
In this setting, the effective action of the FKM model in the region $0<q<R_G$ is expanded by $q$ as 
\begin{multline}
  S_{\rm eff}(q;U) \\
  = -\gamma N_c \Tr\left[\sum_{i=1}^{b_G} q^{l_i} \left(
   \sum_{C_a\in {\cal F}_i} \left(U_{C_a} + U_{C_a}^{-1}\right)
   + \sum_{C'\in {\cal G}_i} \left(U_{C'} + U_{C'
  }^{-1}\right)
  \right) + {\cal O}(q^{l_{b_G}+1})
  \right]\,. 
\end{multline}

When ${\cal G}_i =  \emptyset$ for all $i=1,\cdots,b_G$, the situation is the easiest: 
For large $\gamma$ with $\gamma q^{l_i}={\cal O}(1)$, 
Wilson loops of the length more than $l_i$ are decoupled from the system, 
and the effective action in the limit of $\gamma\to\infty$ can be regarded as a summation of independent GWW models. 
Therefore, we can expect that the GWW phase transitions occur at 
\begin{equation}
  q^*_{i} \simeq (2\gamma)^{1/l_i}\,
  \quad \text{for }i=1,\cdots,b_G.
\end{equation}

On the other hand, if ${\cal G}_i$ for some $i$ is {\it not} empty, the effective action in the same limit is not simply a summation of the GWW models but also includes non-trivial coupling terms of the fundamental Wilson loops. 
Because of the existence of these coupling terms, we cannot use the result of the GWW model. 
In this paper, we examine one of such models numerically in the next section. 
A theoretical analysis of this kind is discussed separately. 

As discussed in the previous section, the FKM model has a dual expression in the region $q>1$ where the effective action can be expanded by $1/q$. 
Therefore, as long as the phase transitions occur in the region $0<q<R_G$, they are expected to occur also in the region $q>1$.

If the FKM model is defined on a regular graph, the region $q>1$ is just a copy of the region $0<q<R_G$ because of the exact duality \eqref{eq:exact duality}. 
However, for an irregular graph, the phase structure in $q>1$ is different from that in $0<q<R_G$ in general. 
In this case, we have to write down the $1/q$-expansion of the effective action \eqref{eq:Seff dual}, which generally takes the form, 
\begin{multline}
  S_{\rm eff}(q;U) \\
  = -\gamma N_c \Tr\left[\sum_{i=1}^{b_G} g_i(q) \left(
   \sum_{C_a\in {\cal F}_i} \left(U_{C_a} + U_{C_a}^{-1}\right)
   + \sum_{C'\in {\cal G}_i} \left(U_{C'} + U_{C'
  }^{-1}\right)
  \right) + {\cal O}(q^{-l_{b_G}-1})
  \right]\,, 
\end{multline}
where $g_i(q)$ is a function of $q$ which is defined by $1/q$-expansion. 
When all ${\cal G}_i$ are empty, we can expect that the phase transitions occur at $\tilde{q}_i^* > 1$ with 
\begin{equation}
  \gamma g_i(\tilde{q}_i^*) = \frac{1}{2}\,,
\end{equation}
as well as in the region $0<q<R_G$. 
On the other hand, if ${\cal G}_i$ is not empty for some $i$, the situation is expected to be different from the GWW model. 
In this case, we have to carry out the numerical computations to see the phase structure.

In summary, the phase structure of the FKM model in large $\gamma$ depends on whether ${\cal G}_i$ are empty or not and whether the graph $G$ is regular or not. 
The cycle graph is a simplest example of the case ${\cal G}_i$ are empty and the graph is regular. 
We will give concretely
an example of a regular graph with non-trivial ${\cal G}_i$ (the tetrahedron) and
examples of irregular graphs with empty ${\cal G}_i$ (the double triangle and the triangle-square) 
in turn. 
We also consider the case that
the number of the phase transitions in $0<q<1$ and $q>1$ is different (the triple triangle).

Before analyzing concrete examples, 
we emphasize that, 
even if we can analyze the FKM model by using the GWW model, the FKM model differs from the GWW model because of the existence of the duality. 
For example, we cannot discuss the phase transition at $q^n=2\gamma-1$, that is, the phase transition in the dual region from the GWW model. 
This is because the theory is defined as a sum of infinitely many Wilson loops.

\subsection{Concrete examples}
\label{subsec:examples}

\subsubsection*{\underline{Tetrahedron}}
\label{subsec:K4}

The tetrahedron ($K_4$) is an example of regular graphs 
as explained in Sec.~\ref{subsec:regular graph} and depicted in Fig.~\ref{K4 and DT graph} (a). 
The maximal convergence radius is determined by the (common) degree of the vertices like $R_{K_4}=1/t=1/2$. 
In particular, $q=R_{K_4}=1/2$ is the only nontrivial real pole of the Ihara zeta function \eqref{DT Ihara}. 
The set of six directed edges is defined by 
\begin{equation}
  E=\left\{e_1,\cdots,e_6\right\}\equiv\left\{
  \langle v_1,v_2 \rangle,
  \langle v_1,v_3 \rangle,
  \langle v_1,v_4 \rangle,
  \langle v_2,v_3 \rangle,
  \langle v_3,v_4 \rangle,
  \langle v_4,v_2 \rangle
  \right\}\,,
\end{equation}
where $V=\{v_1, v_2, v_3, v_4\}$ stands for the set of the vertices depicted in Fig.~\ref{K4 and DT graph} (a). 

The rank of $K_4$ is $r=3$ and the set ${\cal F}$ is not unique since it has four triangles, but only three of them can be fundamental. 
We here choose 
\begin{equation}
  {\cal F} = {\cal F}_1 = \{
  [e_1e_4e_2^{-1}], 
  [e_2e_5e_3^{-1}], 
  [e_3e_6e_1^{-1} ]
  \} 
  \equiv 
  \{[C_1],[C_2],[C_3]\}. 
\end{equation}
Then, the remaining triangle $e_4e_5e_6$ is regarded as a composite cycle,
\begin{equation}
  e_4e_5e_6 = C_1C_2C_3\,.
\end{equation}
Therefore, we see $b_{K_4}=1$ and 
\begin{equation}
  l_1=3, \quad m_1=3, \quad 
  {\cal G}_1 = \{ [C_1C_2C_3] \}\,. 
\end{equation}

The $q$-expansion of the effective action is given by 
\begin{equation}
  S_{\rm eff}(q;U) = -\gamma N_c \Tr\Big[
  q^3 \left(
  U_1 + U_2 + U_3 + U_1U_2U_3 + {\rm h.c.}
  \right)
  + {\cal O}(q^{-4})
  \Big]\,,
  \label{eq:Seff K4 limit}
\end{equation}
where the interaction term $U_1U_2U_3$ exists because ${\cal G}_1$ is not empty, 
which prevents the theory from being associated with the GWW model.
We will investigate the nature of the phase transition numerically in the next section. 
Since $K_4$ is regular, the FKM model on $K_4$ has the exact duality between the regions $0<q<R_{K_4}$ and $q>1$. 
We will also check it numerically in the next section.

\subsubsection*{\underline{Double triangle}}
\label{subsec:DT}

As explained in Sec.~\ref{subsec:irregular graph} and depicted in Fig.~\ref{K4 and DT graph} (b), 
the double triangle (DT) is an example of irregular graphs and 
the maximal convergence radius is given by $R_{\rm DT}=0.65729\cdots$.  
We see that $q=R_{\rm DT}$ is the only non-trivial real pole of $\zeta_{\rm DT}(q)$. 
The set of five directed edges is defined by
\begin{equation}
  E = \left\{e_1,\cdots,e_5\right\}\equiv\left\{
  \langle v_1,v_2 \rangle,
  \langle v_2,v_3 \rangle,
  \langle v_3,v_4 \rangle,
  \langle v_4,v_1 \rangle,
  \langle v_3,v_1 \rangle
  \right\}\,,
\end{equation}
where $V=\{v_1, v_2, v_3, v_4\}$ stands for the set of the vertices depicted in Fig.~\ref{K4 and DT graph} (b). 

The rank of DT is $r=2$ and the set of generators of the minimal lengths is
\begin{equation}
  {\cal F} = {\cal F}_1 = \{[e_1e_2e_5],[e_5^{-1}e_3e_4]\}\,,
\end{equation}
which consists of two triangles. 
Since there is no other triangle in the cycles of DT, we see $b_{\rm DT}=1$ and 
\begin{equation}
  l_1=3, \quad m_1=2, \quad {\cal G}_1=\emptyset\,.
\end{equation}
Correspondingly, the $q$-expansion of the effective action in $0<q<R_{\rm DT}$ is given by 
\begin{equation}
  S_{\rm eff}(q;U) = -\gamma N_c \Tr\left[
  \sum_{a=1,2} q^3 \left(U_a + U_a^{-1}\right)
  + {\cal O}(q^{-4})
  \right]\,.
  \label{eq:Seff DT small q}
\end{equation}
Therefore, we expect that the FKM model on DT exhibits the GWW phase transition only once at 
\begin{equation}
  q^* \simeq (2\gamma)^{-\frac{1}{3}}\,,
\end{equation}
in the region $0<q<R_G$ for large $\gamma$. 

On the other hand, for $q>1$, the $1/q$-expansion of the dual effective action \eqref{eq:Seff dual} on DT is given by
\begin{equation}
  S_{\rm eff}(q;U) = -\gamma N_c \Tr\left[
  g(q) \sum_{a=1,2} \left(U_a + U_a^{-1}\right)
  + {\cal O}(q^{-4})
  \right],
  \label{eq:Seff DT large q}
\end{equation}
up to constant terms, where 
\begin{equation}
  g(q) = \frac{1}{4q^3} - \frac{3}{16q^5} + \frac{17}{64q^7} - \frac{63}{256q^9} + {\cal O}(q^{-11})\,. 
\end{equation}
As expected, it is not completely symmetric to the region $0<q<R_{\rm DT}$,
although there is still a symmetry between $U_1$ and $U_2$. 
The position of the phase transition in the region $q>1$ is expected to be the solution of 
\begin{equation}
  \gamma g(\tilde{q}^*) = \frac{1}{2}\,,
\end{equation}
for large $\gamma$.

\subsubsection*{\underline{Triangle-square}}
\label{subsec:TS}

\begin{figure}[H]
  \begin{center}
  \includegraphics[scale=0.75]{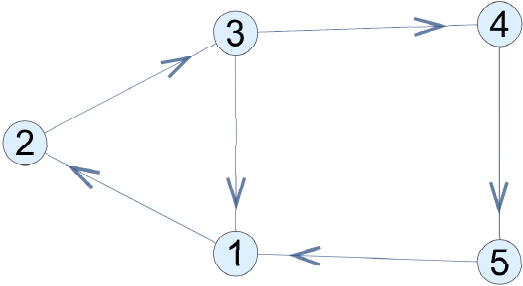}
  \end{center}
  \caption{A triangle-square (TS)}
  \label{TS graph}
  \end{figure}
The triangle-square (TS) is an asymmetric irregular graph which contains two cycles the different length 3 and 4
(See Fig.~\ref{TS graph}).
The Ihara zeta function of TS is given by
\be
\zeta_{\rm TS}(q)
= \frac{1}{(1-q^2)(1+q+q^2-q^4-2q^5)(1-q+q^2-2q^3+q^4-2q^5)}.
\ee
The maximal convergence radius $R_{\rm TS}$ of $\zeta_{\rm TS}(q)$ is read from this zeta function as $R_{\rm TS}=0.70222\cdots$ 
and $q=R_{\rm TS}$ is only the non-trivial real pole of $\zeta_{\rm TS}(q)$. 
The set of six directed edges is defined by
\begin{equation}
  E = 
  \left\{e_1,\cdots,e_6\right\}\equiv\left\{
  \langle v_1,v_2 \rangle,
  \langle v_2,v_3 \rangle,
  \langle v_3,v_4 \rangle,
  \langle v_4,v_5 \rangle,
  \langle v_5,v_1 \rangle,
  \langle v_3,v_1 \rangle
  \right\}\,,
\end{equation}
where $V=\{v_1,v_2, v_3, v_4, v_5\}$ stands for the set of the vertices depicted in Fig.~\ref{TS graph}. 

The rank of TS is again $r=2$ and the set of generators of the minimal lengths is
\begin{equation}
  {\cal F} = \{[e_1e_2e_6],[e_6^{-1}e_3e_4e_5]\}\,,
\end{equation}
which consists of a triangle and a square. 
In this case, ${\cal F}$ is the disjoint union of 
\begin{equation}
  {\cal F}_1 = \{[e_1e_2e_6]\}\quad {\rm and} \quad 
  {\cal F}_2 = \{[e_6^{-1}e_3e_4e_5]\}\,.
\end{equation}
Since there is no other triangle and square in the cycles of TS, we see $b_{\rm TS}=2$ and  
\begin{equation}
   (l_1,l_2)=(3,4), \quad (m_1,m_2)=(1,1),\quad 
   ({\cal G}_1,{\cal G}_2)=(\emptyset,\emptyset)\,.
\end{equation}
Correspondingly, the $q$-expansion of the effective action in $0<q<R_{\rm TS}$ is given by 
\begin{equation}
  S_{\rm eff}(q;U) = -\gamma N_c \Tr\Big[
  q^3 \left(U_1 + U_1^{-1}\right)
  +q^4 \left(U_2 + U_2^{-1}\right)
  + {\cal O}(q^{-5})
  \Big]\,.
  \label{eq:Seff TS}
\end{equation}
Therefore, we expect that the FKM model on TS exhibits the GWW phase transition at 
\begin{equation}
  q_1^* \simeq (2\gamma)^{-\frac{1}{3}}\quad {\rm and}\quad 
  q_2^* \simeq (2\gamma)^{-\frac{1}{4}}\,,
\end{equation}
in the region $0<q<R_G$ for large $\gamma$. 

On the other hand, the $1/q$-expansion of the dual effective action \eqref{eq:Seff dual} in $q>1$ is given by
\begin{equation}
  S_{\rm eff}(q;U) = -\gamma N_c \Tr\left[
  \sum_{a=1,2} g_a(q) \left(U_a + U_a^{-1}\right)
  + {\cal O}(q^{-5})
  \right]
  \label{dual expansion of TS}
\end{equation}
up to constant terms, where 
\begin{equation}
  \begin{split}
  g_1(q) &= \frac{1}{4 q^3} - \frac{3}{16 q^5} + \frac{17}{64 q^7} - \frac{39}{256 q^9} + {\cal O}(q^{-11}), \\
  g_2(q) &= \frac{1}{4 q^4} - \frac{3}{16 q^6} + \frac{17}{64 q^8} - \frac{63}{256 q^{10}} + {\cal O}(q^{-12}). 
  \end{split}
\end{equation}
Again, 
it is not completely symmetric to the region $0<q<R_{\rm TS}$  
although the structure is similar. 
From the effective action, the FKM model on TS is expected to undergo phase transitions twice in the dual region $q>1$ as well as in $0<q<R_{\rm TS}$, 
and the positions of the phase transitions $\tilde{q}_i^*$ in $q>1$ are expected to be the solutions of 
\begin{equation}
  \gamma g_i(\tilde{q}_i^*) = \frac{1}{2}\,,
\end{equation}
for large $\gamma$.

\subsubsection*{\underline{Triple triangle}}
\label{subsec:TT}

\begin{figure}[H]
  \begin{center}
  \includegraphics[scale=0.75]{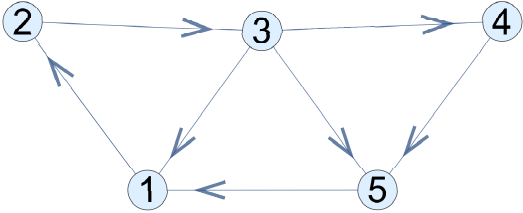}
  \end{center}
  \caption{A triple triangle (TT)}
  \label{TT graph}
\end{figure}

The triple triangle (TT) is an irregular graph which contains three cycles
(See Fig.~\ref{TT graph}).
The Ihara zeta function of TT is given by
\be
\zeta_{\rm TT}(q)
= \frac{1}{\left(2 q^{2}+q+1\right) \left(3 q^{3}+q-1\right) \left(2 q^{4}+q^{3}+2 q^{2}+q+1\right) \left(q+1\right)^{2} \left(q-1\right)^{3}}
\ee
The maximal convergence radius $R_{\rm TT}$ of $\zeta_{\rm TT}(q)$ is read from this zeta function as $R_{\rm TT}=0.53656\cdots$
and $q=R_{\rm TT}$ is the only non-trivial real pole of $\zeta_{\rm TT}(q)$. 
The set of seven directed edges is defined as 
\begin{equation}
  E=\left\{e_1,\cdots,e_7\right\}\equiv\left\{
  \langle v_1,v_2 \rangle,
  \langle v_2,v_3 \rangle,
  \langle v_3,v_4 \rangle,
  \langle v_4,v_5 \rangle,
  \langle v_5,v_1 \rangle,
  \langle v_3,v_1 \rangle,
  \langle v_3,v_5 \rangle
  \right\}\,,
\end{equation}
where $V=\{v_1, v_2, v_3, v_4, v_5\}$ expresses the set of the vertices depicted in Fig.~\ref{TT graph}.

The rank of TT is again $r=3$ and the set of generators of the minimal lengths is
\begin{equation}
  {\cal F} = {\cal F}_1 = \{[e_1e_2e_6],[e_3e_4e_7^{-1}],[e_5e_6^{-1}e_7]\}\,.
\end{equation}
Since there is no other triangle in the cycles of TT, we see $b_{\rm TT}=1$ and 
\begin{equation}
  l_1=3, \quad m_1=3, \quad {\cal G}_1 = \emptyset\,.
\end{equation}
Correspondingly, the $q$-expansion of the effective action in $0<q<R_{\rm TT}$ is given by 
\begin{equation}
  S_{\rm eff}(q;U) = -\gamma N_c \Tr\left[
  \sum_{a=1,2,3} q^3 \left(U_a + U_a^{-1}\right)
  + {\cal O}(q^{-4})
  \right]\,.
  \label{small q expansion of TT}
\end{equation}
Therefore, we expect that the FKM model on TS exhibits the GWW phase transition only once at 
\begin{equation}
  q^* \simeq (2\gamma)^{-\frac{1}{3}}\,,
\end{equation}
in the region $0<q<R_G$ for large $\gamma$.

Interestingly, for $q>1$, the $1/q$-expansion of the dual effective action \eqref{eq:Seff dual} on TT splits as follows
\begin{align}
  S_{\rm eff}(q;U) = -\gamma N_c \Tr\Bigl[
  g_1(q) \left(U_1 + U_2 + U_1^{-1}+U_2^{-1}\right)
  +g_2(q) \left(U_3 + U_3^{-1}\right)
  + {\cal O}(q^{-4})
  \Bigr]
  \label{large q expansion of TT}
\end{align}
up to constant terms, where 
\begin{equation}
  \begin{split}
  g_1(q) &= 
  \frac{1}{6 q^{3}}-\frac{11}{72 q^{5}}+\frac{233}{864 q^{7}} + {\cal O}(q^{-9})\,, \\
  g_2(q) &= \frac{1}{12 q^{3}}-\frac{17}{144 q^{5}}+\frac{115}{576 q^{7}} + {\cal O}(q^{-9}). 
  \end{split}
\end{equation}
Therefore,
as opposed to the cases of DT and TS, 
the FKM model in $q>1$ on TT is not only asymmetric with the region $0<q<R_{\rm TT}$, but also the phase structure in large $\gamma$ differs from $q<1$.
It is expected that the model undergo phase transitions twice in the dual region $q>1$ unlike in $0<q<R_{\rm TT}$, 
and the positions of the phase transitions $\tilde{q}_i^*$ ($i=1,2$) in $q>1$ are expected to be the solutions of 
\begin{equation}
  \gamma g_i(\tilde{q}_i^*) = \frac{1}{2}\,,
\end{equation}
for large $\gamma$. 

\section{Numerical Results I: Regular Graphs}
\label{sec:Regular}

In this section, we numerically examine the phase structure of the FKM model on the regular graphs to check the analytic investigation described so far. 

\subsection{Algorithm and parameter settings}
\label{subsec:algorithm}

The simulations are carried out by using the HMC algorithm. 
The action used in the code is defined through the path representation of the unitary matrix weighted Ihara zeta function \eqref{eq:path Uzeta} to optimize the calculation of the Hamiltonian and the force. 

The observable which we mainly measure is the specific heat, 
which 
shows a cusp behavior at the phase transition point if the transition is of the third-order as the standard of the GWW phase transition. 
We will plot the behavior of the specific heat for each graph described in Sec.~\ref{subsec:examples} toward the parameter $s$ defined in \eqref{eq:q vs s general} or \eqref{eq:s Cn} for $C_n$ in order to see the dual expression discussed in Sec.~\ref{subsec:duality} clearly. 
If the graph is regular, the plot of any observable must be 
fold-symmetric around $s=1/2$, 
while the symmetry must be broken if the graph is irregular.

The regions of the coupling constant $s$ to carry out the simulations are $s>1$ and $s<0$ since the FKM model is unstable in the critical strip $0<s<1$ as discussed in Sec.~\ref{subsec:Stability}. 
Although the simulations fail when $s$ is close to the boundary of the critical strip, $s=0$ and $s=1$, it is natural as discussed at the end of Sec.~\ref{subsec:Stability} and it does not affect to our purpose to see the phase structure of the model.

Since the GWW phase transition takes place in large $N_c$, we have to use sufficiently large $N_c$ to see the nature of the transition. 
On the other hand, the cost of the computation becomes higher when we adopt larger value of $N_c$. 
Therefore, we have to find an appropriate value of $N_c$. 
For this purpose, we use the model on the cycle graph where the exact result is known as shown in Sec.~\ref{subsec:cycle graph}. 
As we will see shortly in Sec.~\ref{subsec:sim Cn}, $N_c=16$ is sufficient to consider $N_c$ to be {\it large}.

The parameters of the molecular dynamics (the number of divisions $N_\tau$ and the time step $D_\tau$) are tuned so that the acceptance ratio exceeds $0.75$ with fixing the combination $N_\tau D_\tau$ to be $0.1$.
We start the simulation with $U_a=\bsone_{N_c}$ $(a=1,\cdots,r)$ and discard the first $50000$ trajectories for thermalization. 
After that, we compute the observables every ten trajectories and have gathered $20000$ samples for each set of parameters.
The average and standard error of each physical quantity are evaluated using the jackknife method with a bin size for which the autocorrelation becomes sufficiently small. Correspondingly, the number of independent samples is $1000$--$10000$ for each parameter.

\subsection{Cycle graph}
\label{subsec:sim Cn}

\begin{figure}[ht]
  \begin{center}
    \includegraphics[scale=0.42]{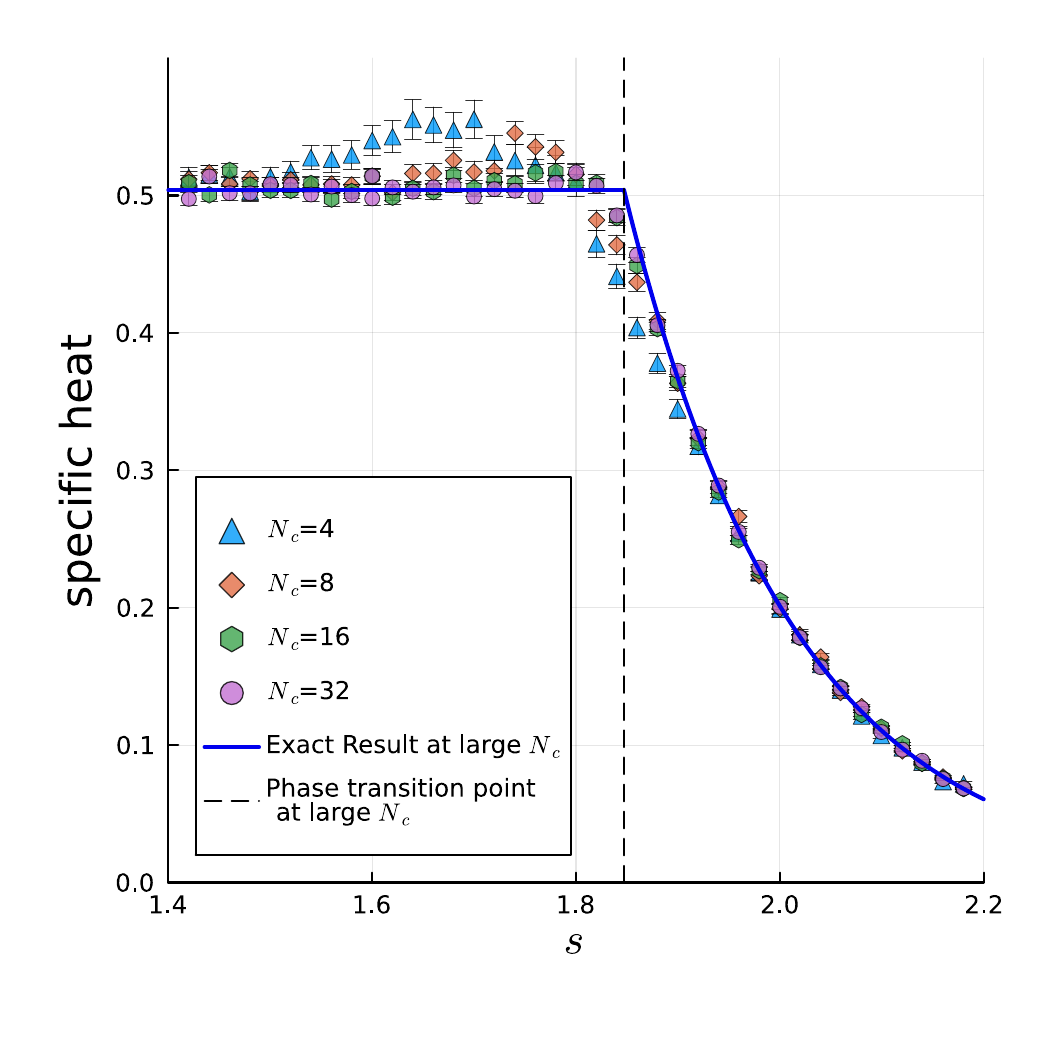}
  \end{center}
  \vspace{-10mm}
  \caption{
  The plots of the specific heat of the FKM model on $C_3$ with varying $N_c=4,8,16,32$ and fixed $\gamma=128$.
  The solid line indicates the exact result obtained in the large $N_c$ limit.
  The vertical dashed line shows the phase transition point which is also evaluated at large $N_c$.}
  \label{C3 specific heat with various Nc}
\end{figure}

The cycle graph $C_n$ is a 2-regular graph with $n$ vertices. (See Fig.~\ref{C3 graph}.) 
In this case, we can choose only one representative
unitary matrix $U$ after fixing the gauge.
First of all, we simulate and measure the specific heat 
at $\gamma=128$ while varying $N_c$.
The plot of the specific heat near the phase transition point is shown in Fig.~\ref{C3 specific heat with various Nc}. 
The plot indicates that the behavior at $N_c=16$ and $N_c=32$ is almost the same. 
This means that the analytical results in the large $N_c$ limit
are well reproduced even at $N_c=16$.%
\footnote{
On the cycle graph, the autocorrelation at smaller $N_c=4$ is stronger than at $N_c=16$, and the errors of the measurements become large.
In this sense as well, $N_c=16$ is the best choice in the following measurements.
}

\begin{figure}[H]
  \begin{center}
  \includegraphics[scale=0.45]{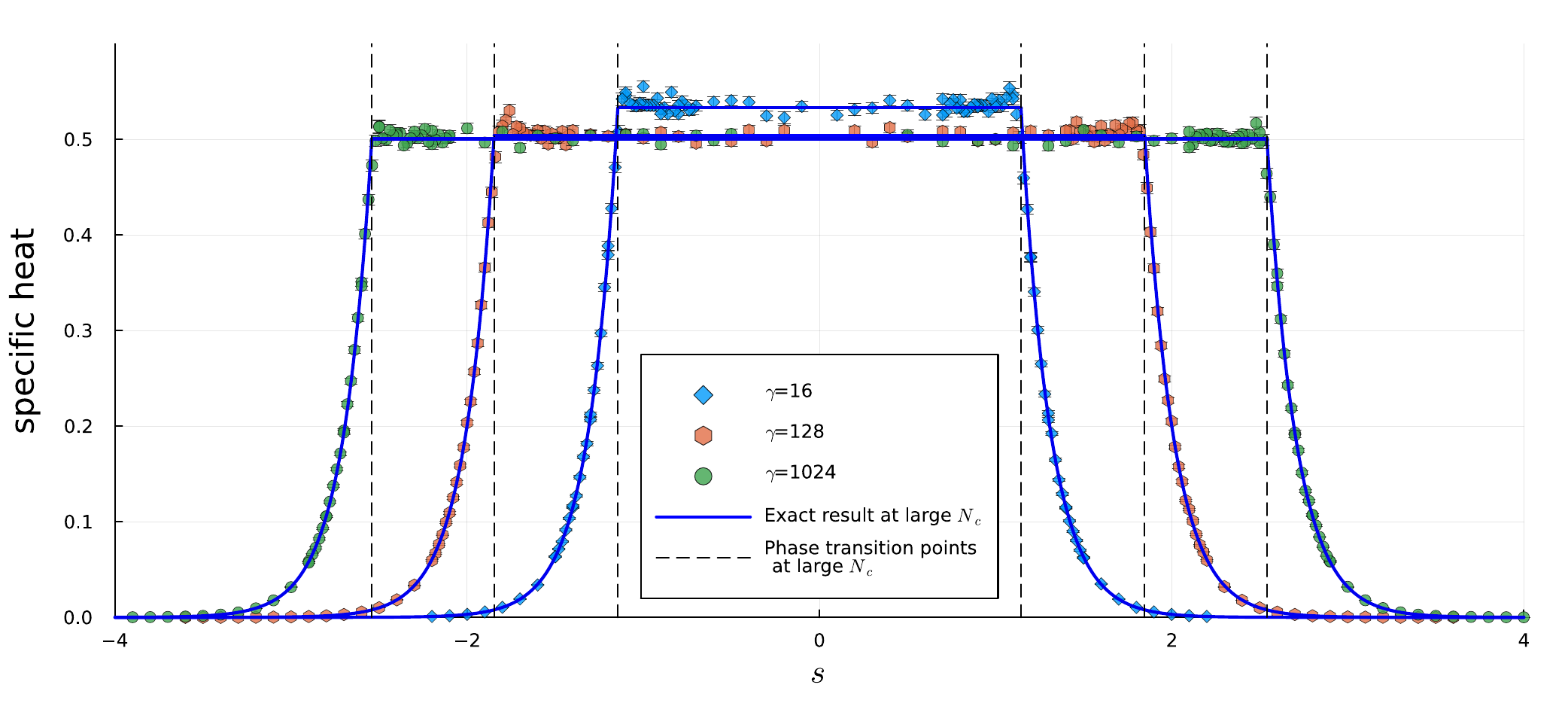}
  \end{center}
  \vspace{-5mm}
  \caption{The numerical results of the specific heat on the $C_3$ graph.
  We have changed $\gamma$ as $\gamma=16,128,1024$ with fixing $N_c=16$.
  The solid line and the vertical dashed line show the analytic result of the specific heat and the phase transition points in the large $N_c$ limit, respectively. 
  }
  \label{C3 specific heat}
  \end{figure}

We then measure the specific heat with fixing $N_c=16$ while varying 
$\gamma=16,128$ and $1024$, 
which is plotted in Fig.~\ref{C3 specific heat} on $s$-axis.
Note that we have removed some statistical outliers of the measurements near
$s=0$, namely the shrunk critical strip, where the numerical simulation
becomes unstable.
The exact solution in the large $N_c$ limit is again drawn by the solid line,
which is valid at any value of finite $\gamma$. 
It shows that the numerical results agree with the analytic results
including the dependence of $\gamma$.
Similarly, it can be seen that the phase transition points of the exact solutions, represented by the dashed lines in the same figure, 
also perfectly match the results of the simulation.
We can also see the exact duality
between the small ($s>1$) and large ($s<0$) coupling regions
as discussed in Sec.~\ref{subsec:cycle graph}.

\begin{figure}[H]
  \begin{center}
  \subcaptionbox{$s=2.15848$}[.32\textwidth]{
  \includegraphics[scale=0.3]{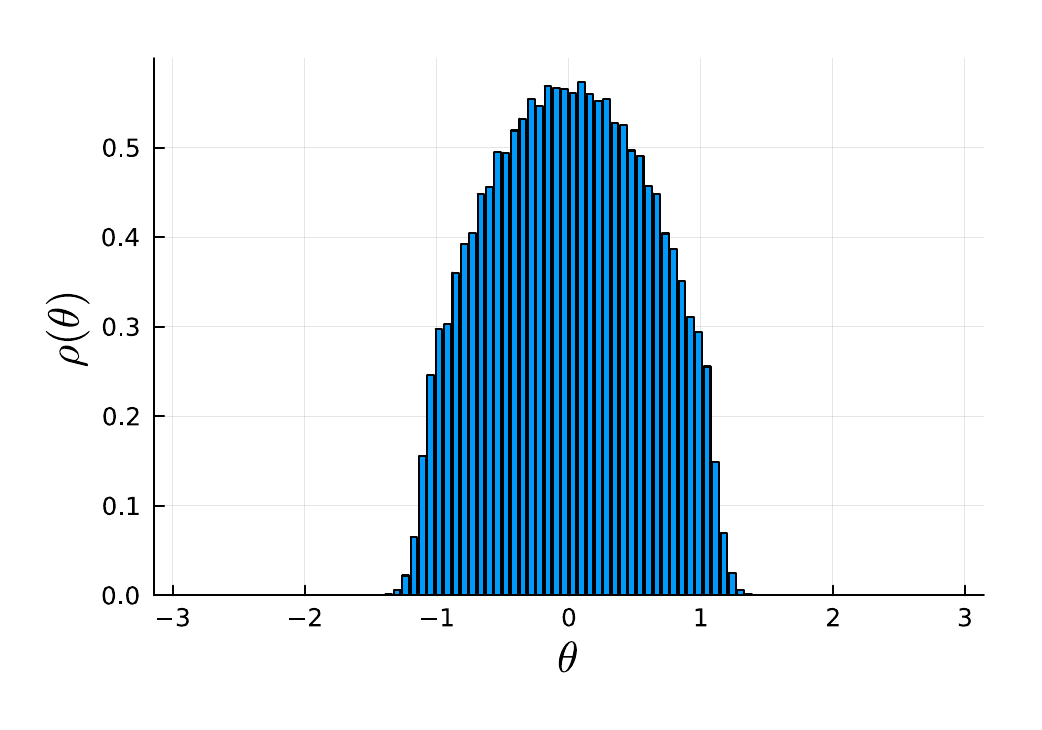}
  }
  \subcaptionbox{$s=2.54084\simeq s^*$}[.32\textwidth]{
  \includegraphics[scale=0.3]{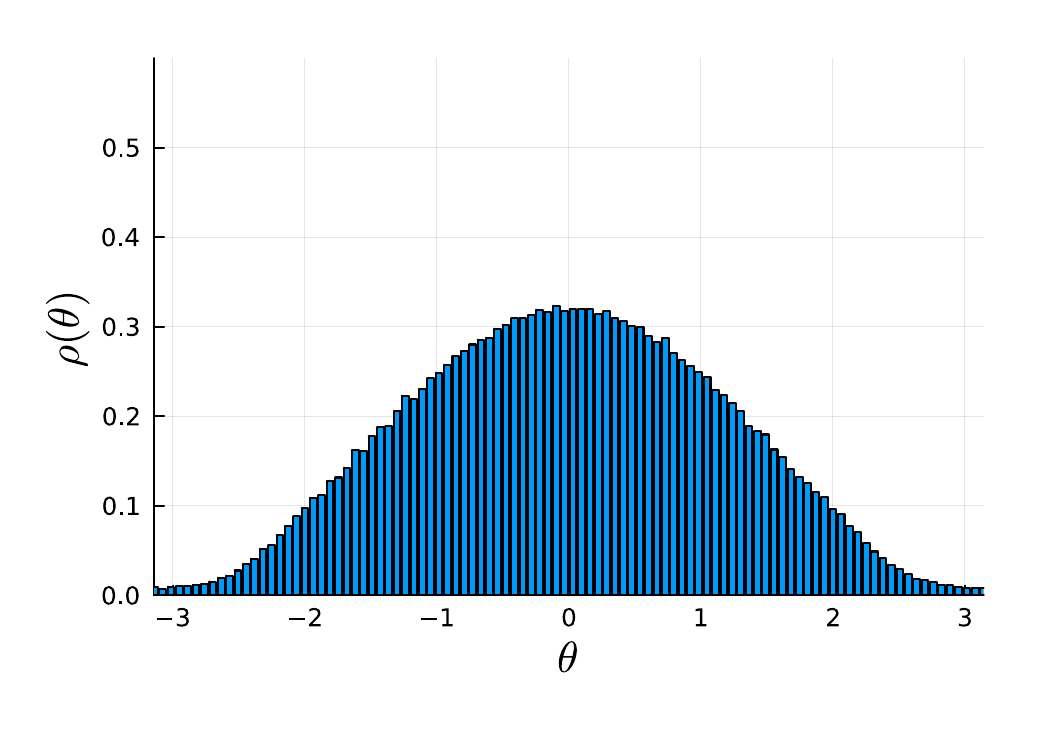}
  }
  \subcaptionbox{$s=2.84215$}[.32\textwidth]{
  \includegraphics[scale=0.3]{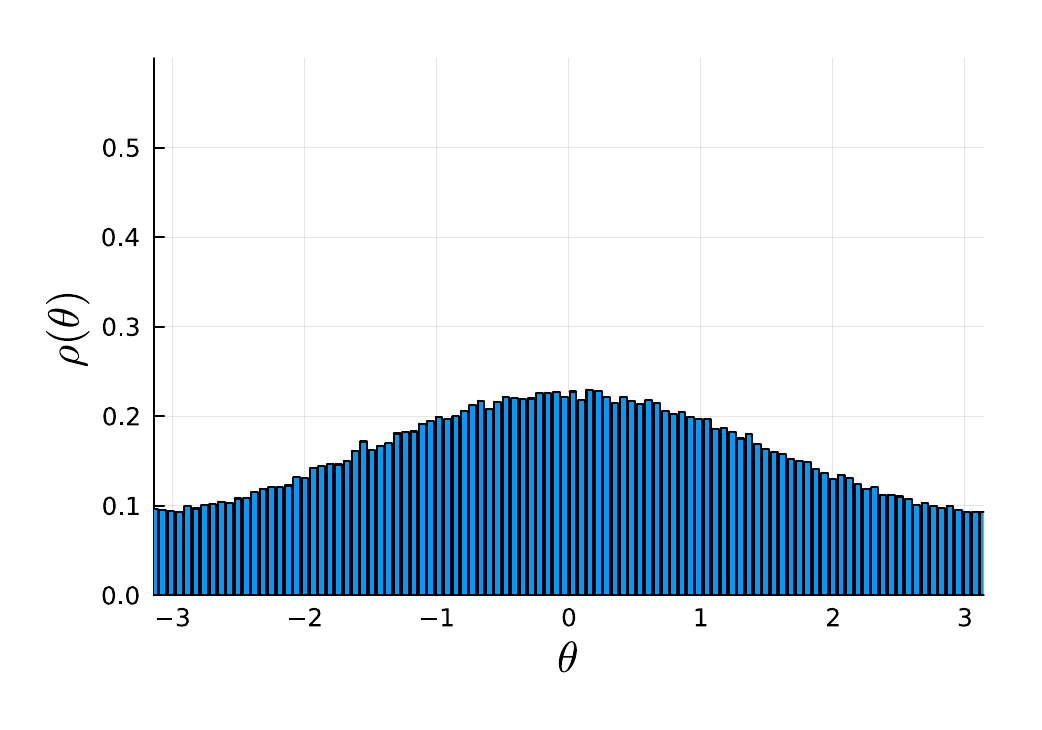}
  }
  \end{center}
  \caption{The eigenvalue distributions of $U$ for $N_c=16$ and $\gamma=1024$.
  (a) and (c) corresponds to the deconfinement and confinement phase, respectively. (b) is the eigenvalue distribution near the phase transition point.} 
  \label{eigenvalue distribution of C3}
\end{figure}

Fig.~\ref{eigenvalue distribution of C3} shows the eigenvalue distributions of $U$ at three typical parameters of $s$. 
The eigenvalue distribution 
at $s \simeq s^*$ is depicted 
in Fig.~\ref{eigenvalue distribution of C3} (b), and those of smaller and larger values of $s^*$ are depicted in (a) and (c), respectively.
This result 
indicates that this phase transition is the GWW phase transition. 
In other words, it separates the deconfinement phase (a) and
the confinement phase (c). 

The results of the above numerical calculations perfectly reproduce the analytical results in the large $N_c$ limit, which means that our simulation code can be trusted.

\subsection{Tetrahedron}

The tetrahedron, considering as a graph, is the complete graph $K_4$ depicted in Fig.~\ref{K4 and DT graph} (a).
Since $K_4$ is a 3-regular graph, the FKM model on $K_4$ should have the duality between $q$ and $1/(2q)$ as discussed in Sec.~\ref{subsec:K4}. 
Thus, if we define $s$ by $q= 2^{-s}$ as \eqref{eq:q vs s}, the specific heat must indicate the duality between $s$ and $1-s$ across the critical strip.

As discussed in Sec.~\ref{subsec:K4}, 
we expect that the phase transition occurs 
at the same time for the three independent unitary matrices $U_1$, $U_2$, $U_3$ because of the symmetry of each cycle (face).
Thus, the FKM model on $K_4$ expresses the phase transition only once, both in the small and large $q$ regions.

\begin{figure}[ht]
  \begin{center}
  \includegraphics[scale=0.45]{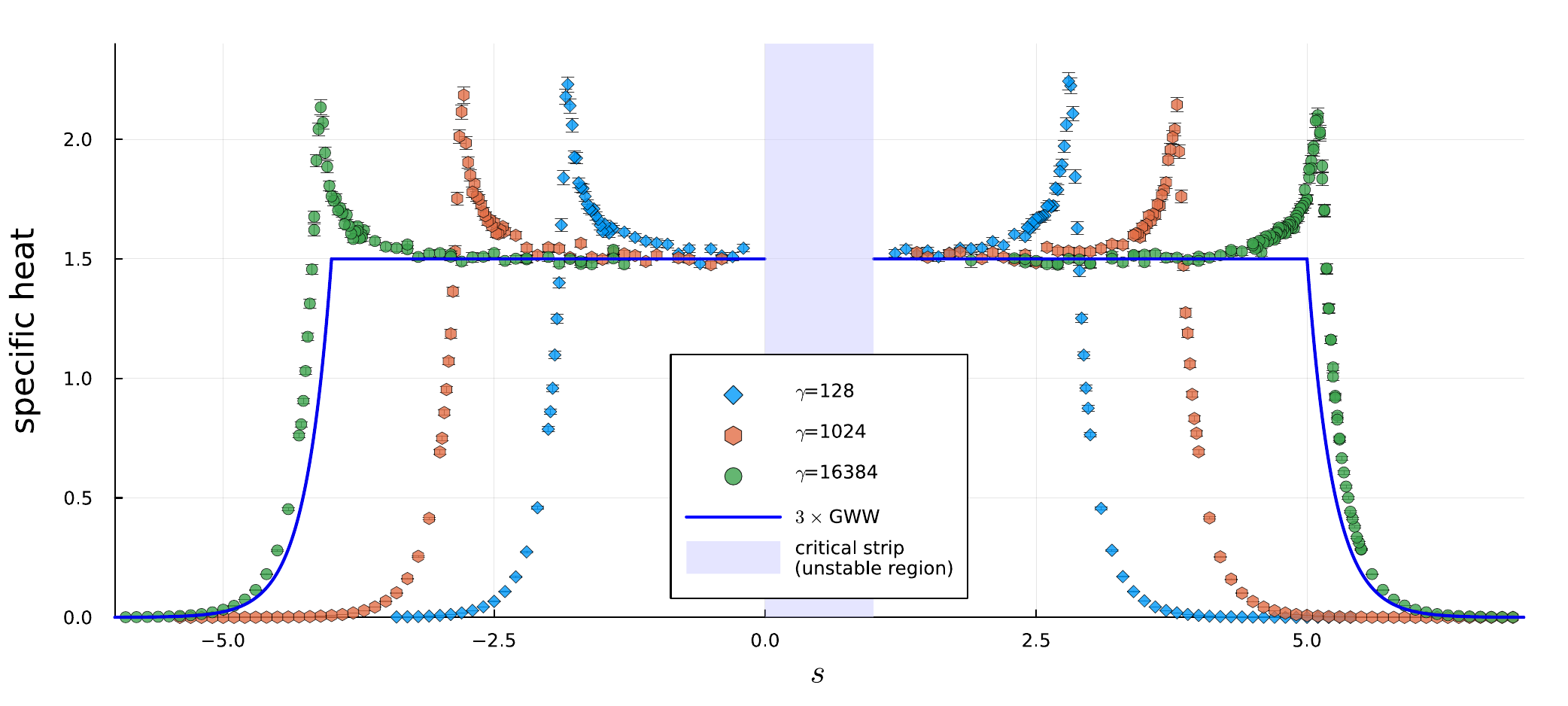}
  \end{center}
  \vspace{-5mm}
  \caption{The specific heat of the FKM model on the tetrahedron $K_4$
  with varying 
  $\gamma=128, 1024, 16384$ and fixing $N_c=16$.
  The solid line, drawn to compare the results of the FKM and GWW models, represents three times the specific heat of the GWW model for $\gamma=16384$.
  }
  \label{Tetrahedron specific heat}
  \end{figure}
The numerical result of the specific heat on $K_4$
is plotted in Fig.~\ref{Tetrahedron specific heat}.
Exact duality is observed between the small ($s>1$) and large ($s<0$) coupling regions, with one phase transition point in each region being perfectly symmetrical.
We also see that the behavior of the specific heat seems to be quite different from
that of the naive GWW approximation as expected from the discussion in Sec.~\ref{subsec:K4}. 
Indeed, comparing the results of the simulation with the solid line in Fig.~\ref{Tetrahedron specific heat}), which is the specific heat of the GWW model for the three independent cycles $U_1$, $U_2$, $U_3$, we see a large discrepancies in the shapes. 
Concretely speaking, the transition points are different and ``pointed elbow'' behaviors can be seen at the phase transition points in the numerical simulation.
This is because the interaction terms $\Tr U_1 U_2 U_3 + \rm{h.c.}$ in \eqref{eq:Seff K4 limit}, which come from the fourth Wilson loop on $K_4$, cannot be negligible even in the large $\gamma$ limit. 
We emphasize that the solid line in Fig.~\ref{Tetrahedron specific heat} is just a sum of the three cycles without the interaction.

\begin{figure}[t]
  \begin{center}
  \subcaptionbox{$s=5.0$}[.45\textwidth]{
  \includegraphics[scale=0.45]{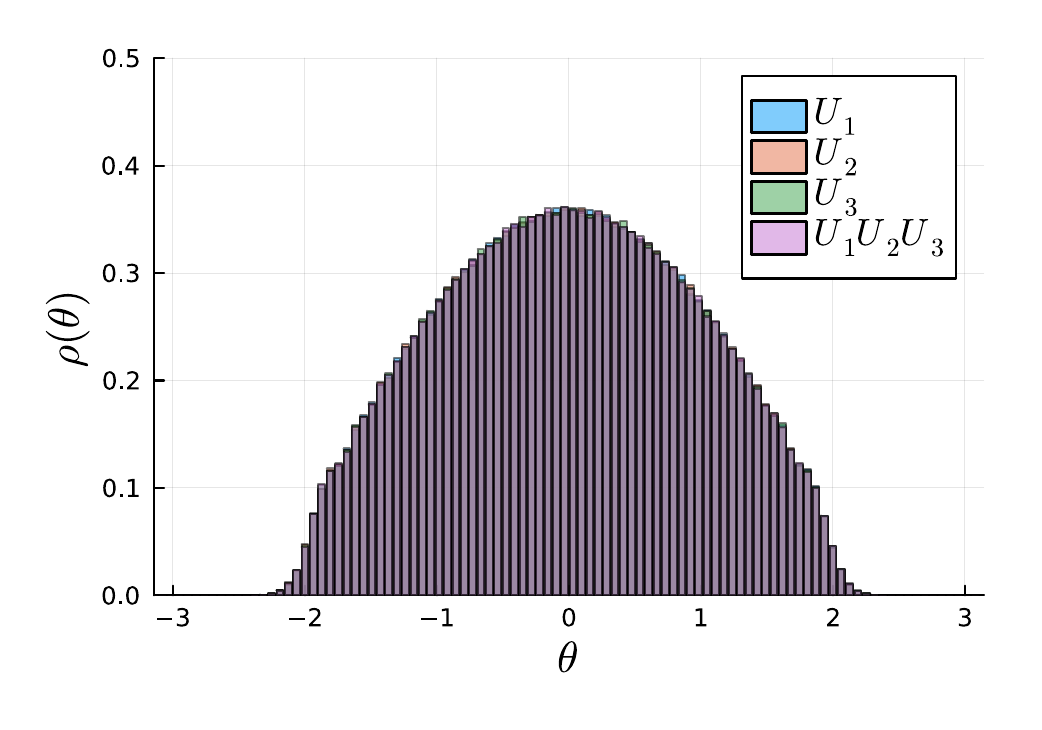}
  }
  \hspace*{0.5cm}
  \subcaptionbox{$s=5.11979$}[.45\textwidth]{
  \includegraphics[scale=0.45]{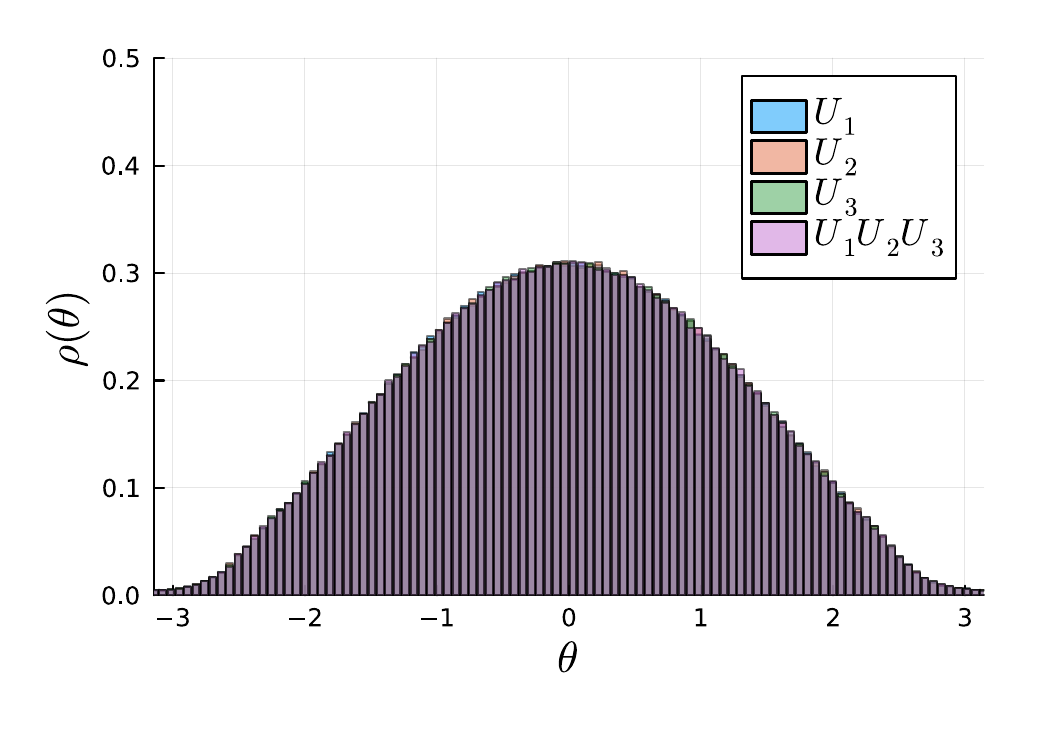}
  }
  \end{center}
  \vspace{-5mm}
  \caption{
  The eigenvalue distribution of $U_1$, $U_2$, $U_3$ and $U_1U_2U_3$ in the FKM model on $K_4$ at $s=5.0$, the phase transition point of the GWW model (a), and at $s=5.11979$, the value of $s$ closest to the peak of the pointed elbow of the specific heat in Fig.~\ref{Tetrahedron specific heat} (b).
  }
  \label{ed for K4}
\end{figure}

To check if the phase transition point of the FKM model is different from that of the GWW model from another point of view, we compare the eigenvalue distributions at different values of $s$ in Fig.~\ref{ed for K4}. 
Ignoring the interaction term allows the application of the GWW model's results, predicting the phase transition point at $s=5.0$. However, the eigenvalue distribution at this value, shown in Fig.~\ref{ed for K4} (a), does not exhibit the characteristics of the GWW phase transition.
On the other hand, 
\setlength{\intextsep}{-5mm}
\begin{wrapfigure}{r}[0pt]{0.50\textwidth}
  \begin{center}
    \includegraphics[scale=0.40]{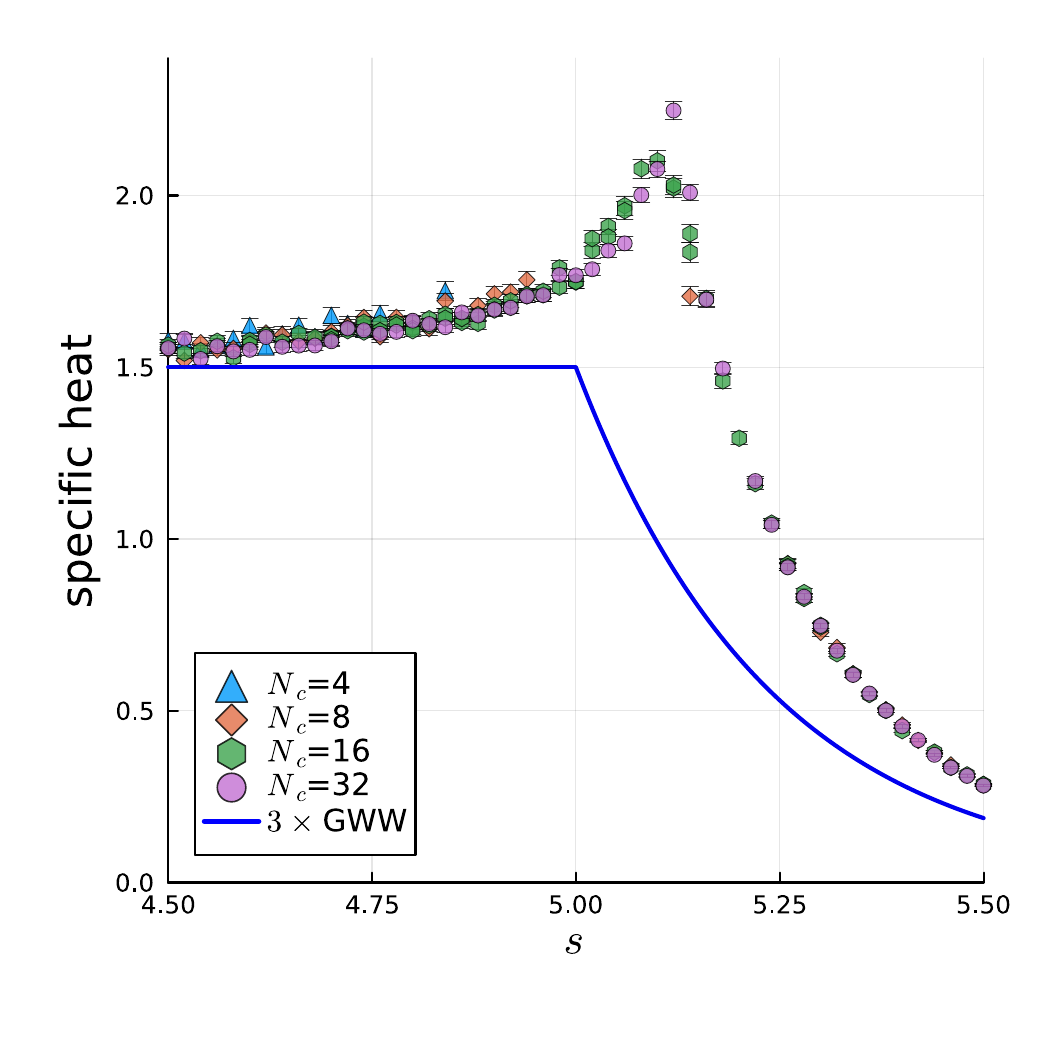}
  \end{center}
  \vspace{-10mm}
  \caption{
  The specific heat of the FKM model on $K_4$
  near the phase transition point with varying $N_c=4,8,16,32$ and fixing $\gamma=16384$.
  }
  \label{Tetrahedron specific heat at various N}
  \vspace{5mm}
\end{wrapfigure}
$s=5.11979$ is the value in our data, 
which is closest to the peak of the pointed elbow in Fig.~\ref{Tetrahedron specific heat},  
and the eigenvalue distribution at this value of $s$, shown in Fig.~\ref{ed for K4} (b), displays the typical shape of the GWW phase transition.
This result confirms that the GWW phase transition occurs at the pointed elbow of Fig.~\ref{Tetrahedron specific heat}, which is shifted from the GWW model by the effect of the interection term.

We next investigate in detail the shape of the ``elbow'' near the phase transition.
Fig.~\ref{Tetrahedron specific heat at various N} shows an enlarged view of the behavior of the specific heat near the phase transition at large $\gamma=16384$.
In order to see the change of the width of the elbow in taking 
the large $N_c$ limit, $N_c=32$ is included.
As shown in the figure, the shape of the elbow does not change when $N_c$ is enlarged. 
This implies that the behavior does not indicate a jump in the specific heat but rather a discontinuous change in the differential coefficient, similar to what is observed in the case of the circle graph.
Therefore, it is reasonable to assume that the order of the phase transition of the FKM model on $K_4$ is again third-order. 
Analytically exploring the significance of the pointed elbow and how the phase transition point shifts due to the interaction term of three cycles is intriguing. These aspects will be discussed separately.

\section{Numerical Results II: Irregular Graphs}
\label{sec:Irregular}

In this section, we numerically examine the phase structure of the FKM model on the irregular graphs, whose zeta functions do not possess exact duality. 
However, we find that the behavior of the specific heat still has a similar structure between the small and large coupling regions.

\subsection{Double triangle}

\vspace{10mm}
\begin{figure}[H]
\begin{center}
\includegraphics[scale=0.45]{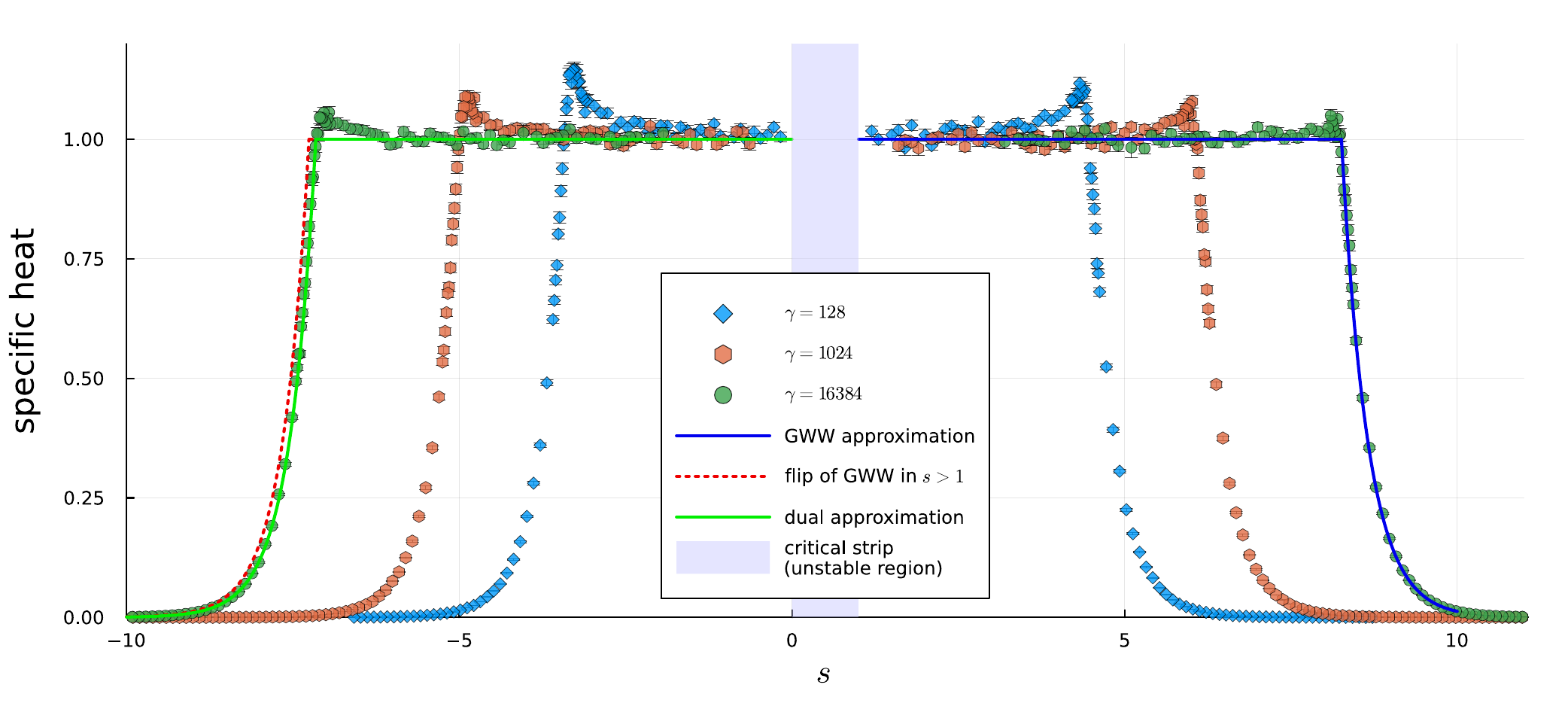}
\end{center}
\vspace{-5mm}
\caption{The specific heat of the FKM model on DT 
with varying $\gamma=128, 1024, 16384$ and fixing $N_c=16$.
The solid lines in $s>1$ and $s<0$ represent the predictions from the effective actions \eqref{eq:Seff DT small q} and \eqref{eq:Seff DT large q} for $\gamma=16384$, respectively.
The dashed line in $s<0$ region is just a flip of the prediction of the GWW model in $s>1$.
}
\label{Double triangle specific heat}
\end{figure}
\vspace{10mm}

The double triangle (DT) is 
depicted in Fig.~\ref{K4 and DT graph} (b). 
As discussed in Sec.~\ref{subsec:irregular graph}, if we define $q=(R_{\rm DT})^s$ with $R_{\rm DT}=0.65729\cdots$,
the critical strip appears at $0<\Real s < 1$.
Since the model is unstable in the critical strip, 
we need to avoid this region in the numerical simulation.

The numerical result of the specific heat is shown in Fig.~\ref{Double triangle specific heat}.
The solid line in the region $s>1$ expresses the specific heat derived from the effective action \eqref{eq:Seff DT small q} in the small $q$ region. 
It well fits the numerical results. 
At first glance, one would see that the duality also holds in DT, 
but it is not correct. 
Indeed, if we simply flip the prediction of the GWW model 
from the small coupling region ($s>1$)
to the large coupling region ($s<0$) as drawn in dashed line, 
the numerical results in $s<0$ are slightly different from it. 
The specific heat derived from the effective action \eqref{eq:Seff DT large q} in the large $q$ region
which is shown as the solid line in $s<0$ in Fig.~\ref{Double triangle specific heat}. 
It well agrees with the numerical results.

Thus, we can conclude that the duality does not exactly hold and 
the asymptotic expansion in the dual side is sufficiently accurate
to explain the numerical result in the FKM model on DT.

\subsection{Triangle-square}

\hspace{10mm}
\begin{figure}[H]
\begin{center}
\includegraphics[scale=0.42]{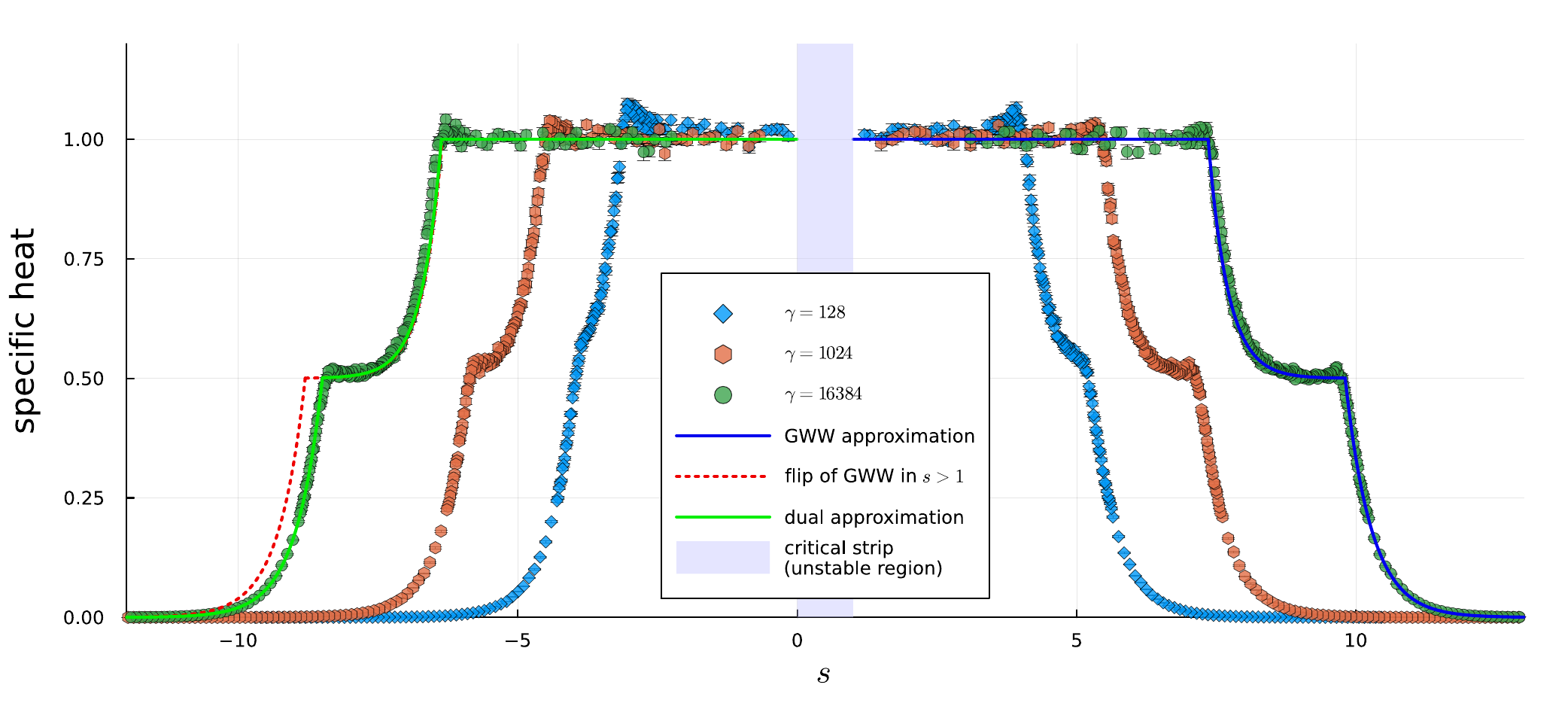}
\end{center}
\vspace{-5mm}
\caption{The specific heat of the FKM model on TS with 
varying $\gamma=128, 1024, 16384$ and fixing $N_c=16$.
The solid lines in $s>1$ and $s<0$ represent the predictions from the effective actions \eqref{eq:Seff TS} and \eqref{dual expansion of TS} for $\gamma=16384$, respectively.
The dashed line in $s<0$ region is just a flip of the prediction of the GWW model in $s>1$.
}
\label{Triangle-square specific heat}
\end{figure}
\vspace{10mm}
The triangle-square (TS) contains two fundamental cycles of different lengths three and four (See Fig.~\ref{TS graph}). 
We plot the specific heat of the FKM model on TS in Fig.~\ref{Triangle-square specific heat}.
Due to the difference of the length of two fundamental cycles, 
two phase transitions appear in small $q$ region ($s>1$). 
The numerical results in $s>1$ well fit to the prediction of the sum of the GWW models of two different coupling constants \eqref{eq:Seff TS}, 
which is depicted by a solid line, as expected.

The duality between the small and large $q$ regions is broken as well as the DT case. 
Indeed, if we flip the prediction of the GWW in the small $q$ region to the large $q$ region as drawn in dashed line, it slightly does not agree with the numerical results in $s<0$.
This slight difference comes from the sub-leading terms of the effective action in the large $q$ region \eqref{dual expansion of TS}.  
The numerical results rather agree with the specific heat predicted by \eqref{dual expansion of TS} drawn in a solid line in the figure. 

\vspace{10mm}
\begin{figure}[ht]
  \begin{center}
  \subcaptionbox{$s=7.35277$}[.45\textwidth]{
  \includegraphics[scale=0.45]{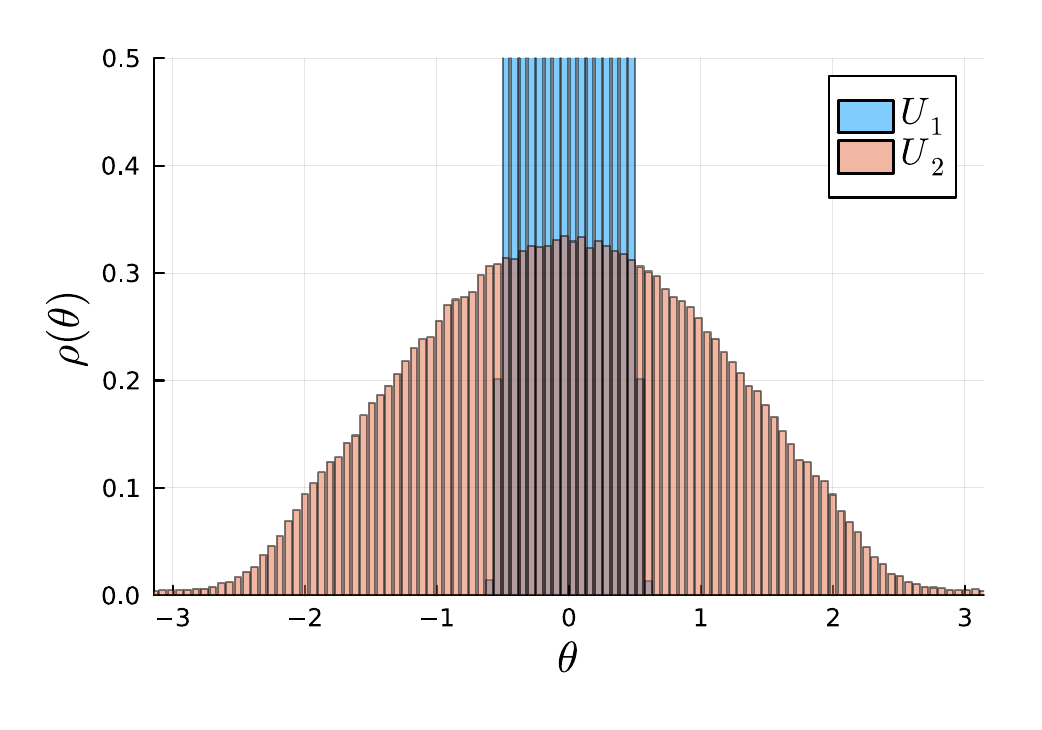}
  }
  \hspace*{0.5cm}
  \subcaptionbox{$s=9.79308$}[.45\textwidth]{
  \includegraphics[scale=0.45]{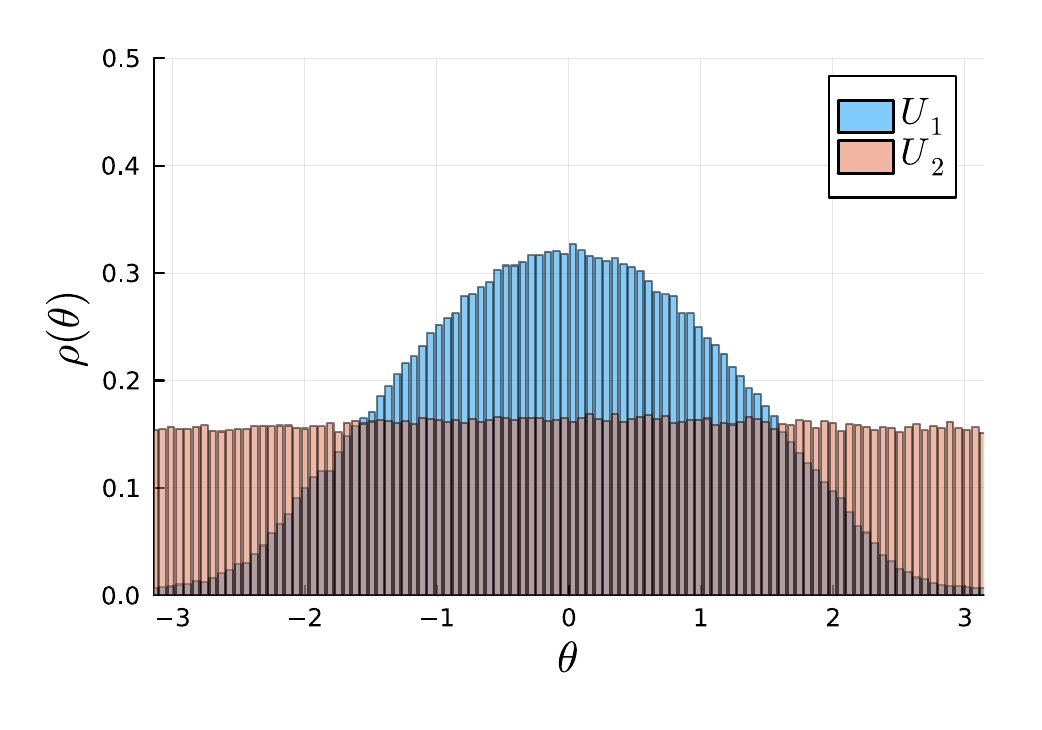}
  }
  \end{center}
  \vspace{-5mm}
  \caption{The eigenvalue distributions of $U_1$ and $U_2$ 
  at the phase transition points of $N_c=16$ and $\gamma=16384$, 
  which are associated with the fundamental cycles of TS with length three and four, respectively. 
  The eigenvalue distribution of $U_1$ in (a) saturates the upper limit of the figure 
  since it is still too sharp and narrow in the deeper deconfinement phase.} 
  \label{ED of TS}
\end{figure}
\vspace{10mm}
The exsistence of the two different phase transitions 
can be also seen from the eigenvalue distributions of the fundamental cycles.
Fig.~\ref{ED of TS} is the eigenvalue distributions of $U_1$ and $U_2$,
which associated with the fundamental cycles of length three and four, respectively, 
at the two phase transition points in $s>1$ at $N_c=16$ and $\gamma=16384$. 
We see that
the eigenvalue distribution of $U_2$ exhibits the GWW phase transition at
$s=7.35277$, while $U_1$ is still in the deconfinement phase.
At the second phase transition point of $s=9.79308$,
the eigenvalue distribution of $U_1$
exhibits the transition. 
Therefore, there exists a mixed phase of the deconfinement and confinement
between $s=7.35277$ and $s=9.79308$.

\subsection{Triple triangle}

\vspace{5mm}
\begin{figure}[H]
\begin{center}
  \includegraphics[scale=0.42]{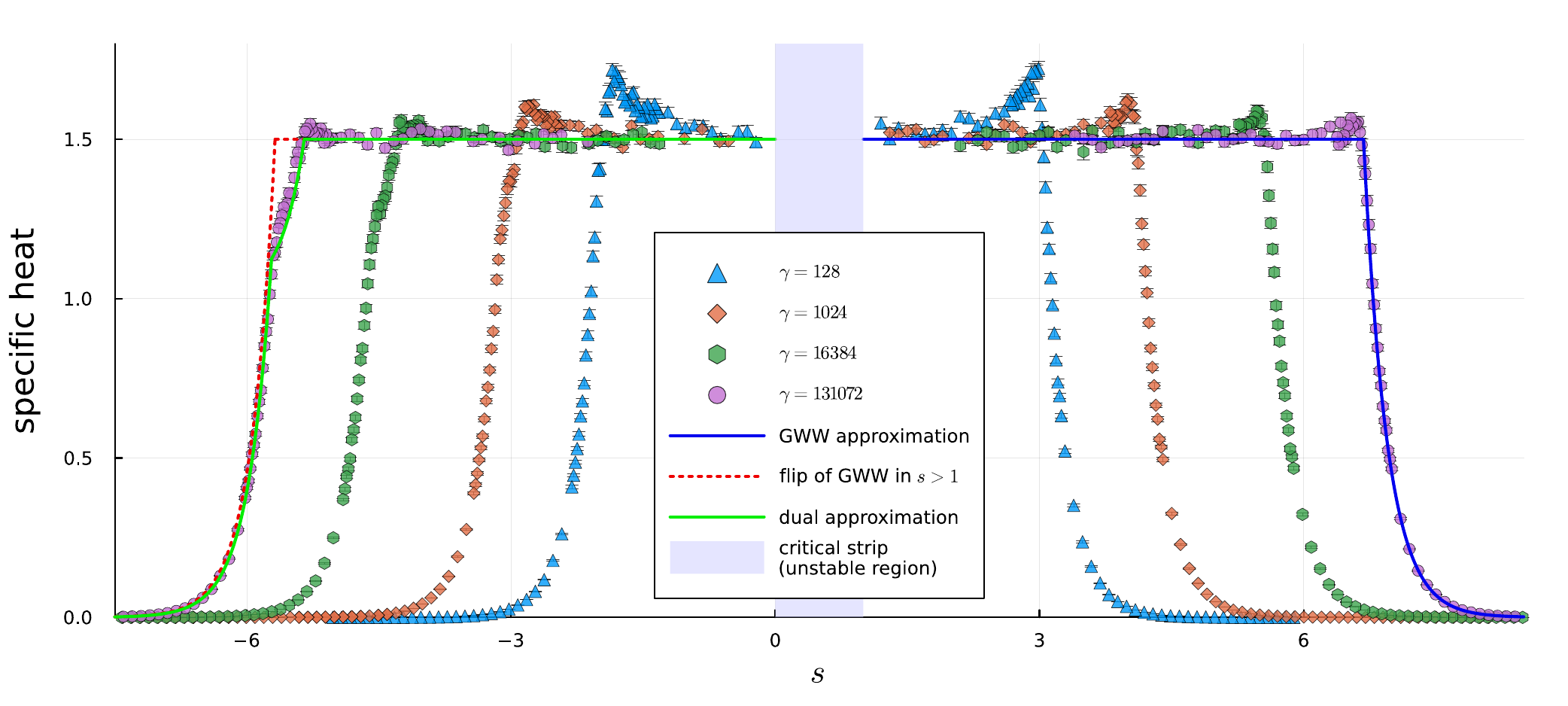}
\end{center}
\vspace{-5mm}
\caption{The specific heat of the FKM model on TT with varying $\gamma=128,1024,16384,131072$ and fixing $N_c=16$. 
The solid lines in the regions $s>1$ and $s<0$ are the predictions from the effective actions 
\eqref{small q expansion of TT} and \eqref{large q expansion of TT}, 
respectively.
The dashed line in the region $s<0$ is just a reflection of the solid line in $s>1$.
}
\label{Triple-Triangle specific heat}
\end{figure}
\vspace{10mm}

\setlength{\intextsep}{-5mm}
\begin{wrapfigure}{r}[0pt]{0.50\textwidth}
  \begin{center}
    \includegraphics[scale=0.45]{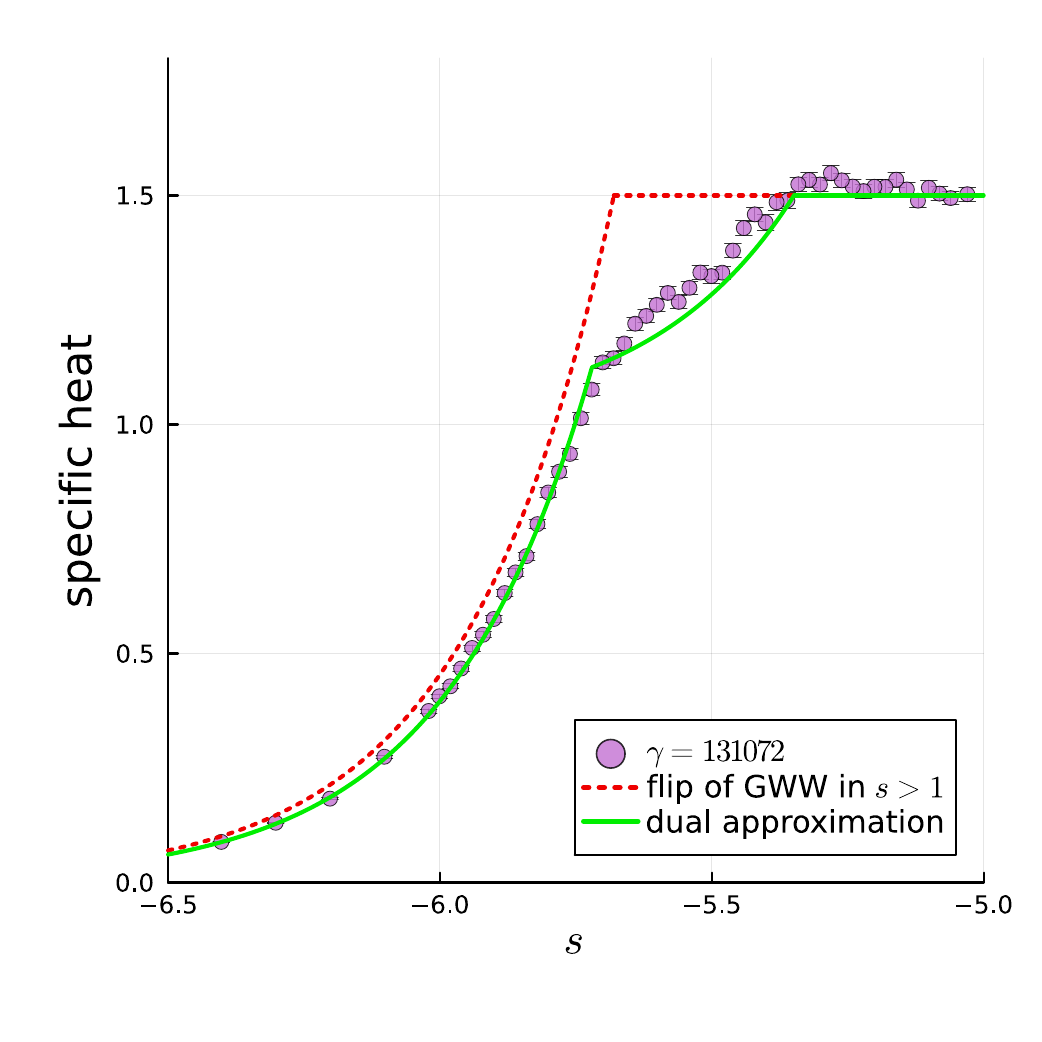}
  \end{center}
  \vspace{-10mm}
  \caption{An enlarged view near the phase transition points for $s<0$ shown in Fig.~\ref{Triple-Triangle specific heat}.}
  \label{Triple-Triangle specific heat magnified}
  \vspace{5mm}
\end{wrapfigure}

We plot the specific heat of the FKM model on triple triangle (TT) in Fig.~\ref{Triple-Triangle specific heat}.
As depicted in Fig.~\ref{TT graph}, TT consists of three triangles connected together. 
Contrary to its apparent simplicity, the FKM model on TT exhibits an interesting behavior.

In the small coupling region $(s>1)$, we see only one phase transition and the numerical results well fit the prediction of the effective action \eqref{small q expansion of TT} drawn by the solid line in $s>1$. 
On the other hand, it is observed that there are two phase transitions in the large coupling region $(s<0)$.
This is expected from the $1/q$ expansion of the effective action in the large coupling reagion \eqref{large q expansion of TT}. 
The small and large coupling regions are obviously not dual.

The difference of the two transition points in $s<0$ is smaller than that of TS shown in Fig.~\ref{Triangle-square specific heat}.
This is due to the fact that, 
in the effective action of TT, 
the phase transition points are separated not by the difference in the powers of $q$ in the coupling constants of the two GWW models but just by the difference in the coefficients, 
while, in the effective action of TS, they separate by the difference in the powers of $q$. 
Since the coupling constants of the GWW models that appear in the effective action are proportional to $\gamma$, we need large $\gamma$ to clearly see the separation of phase transition points in the large $q$ region of TT.
This is why we have added the results of $\gamma=131072$ in Fig.~\ref{Triangle-square specific heat}. 
To see the separation of the phase transition points clearly, we zoom in around the two phase transition points at $s<0$ in Fig.~\ref{Triple-Triangle specific heat magnified}.
From this figure, we observe that the numerical results for $s<0$ do not align with the dashed line, which reflects the solid line for $s>1$. Instead, they align with the solid line, representing the effective action's prediction in the large $q$ region as shown in equation \eqref{large q expansion of TT}.

\vspace{10mm}
\begin{figure}[H]
  \begin{center}
  \subcaptionbox{$s= 6.67970$}[.32\textwidth]{
  \includegraphics[scale=0.3]{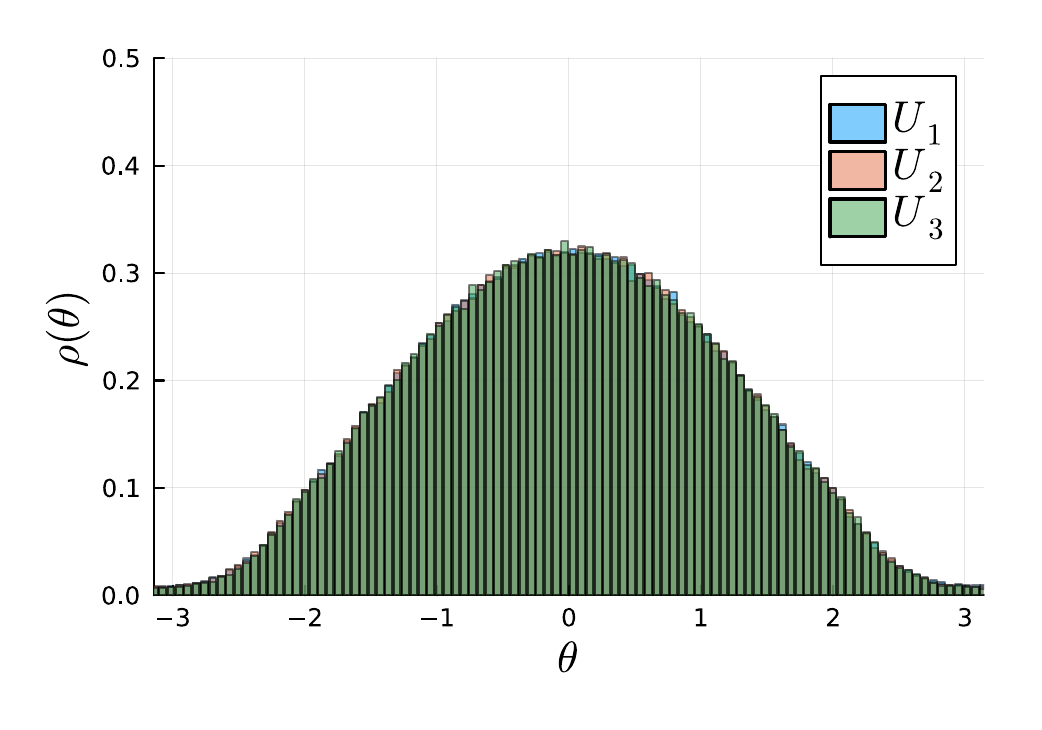}
  }
  \subcaptionbox{$s= -5.72085$}[.32\textwidth]{
  \includegraphics[scale=0.3]{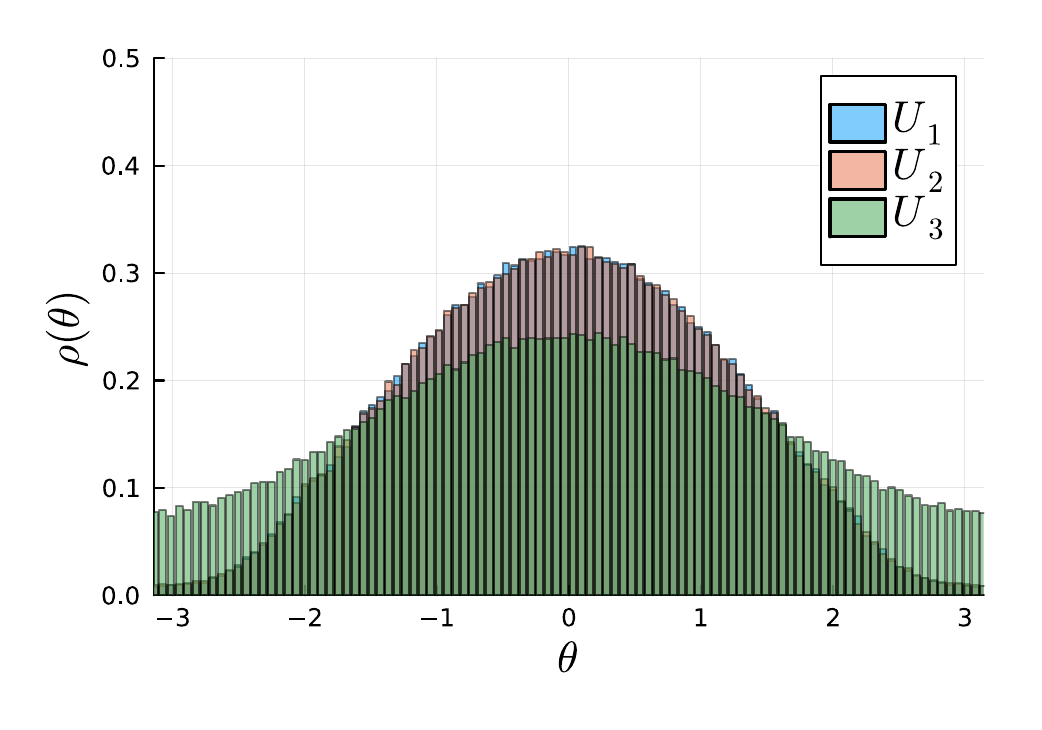}
  }
  \subcaptionbox{$s= -5.34085$}[.32\textwidth]{
  \includegraphics[scale=0.3]{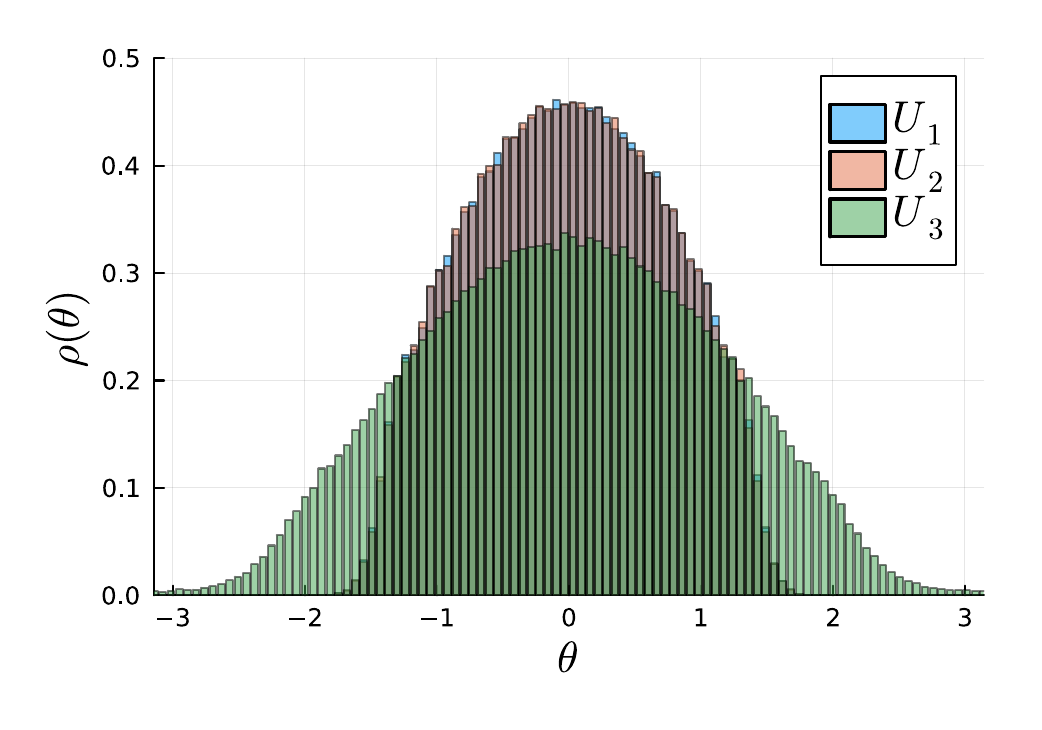}
  }
  \end{center}
  \vspace{-5mm}
  \caption{The eigenvalue distributions of the FKM model on TT with $N_c=16$ and $\gamma=131072$.
  (a) The common phase transition point for all unitary matrices in $s>1$. 
  (b) The phase transition point of $U_1$ and $U_2$ in $s<0$. 
  (c) The phase transition point of $U_3$ in $s<0$. }
  \label{eigenvalue distribution of TT}
\end{figure}
\vspace{10mm}
Fig.\ref{eigenvalue distribution of TT} shows the eigenvalue distributions of the fundamental Wilson loops $U_1$, $U_2$ and $U_3$ with $N_c=16$ and $\gamma=131072$. 
In the small coupling region at $s=6.67970$ shown in (a), all the fundamental Wilson loops exhibits the GWW phase transition at the same time, 
while, in the large coupling region, only $U_3$ behaves differently from $U_1$ and $U_2$, resulting in two different phase transition points as shown in (b) and (c).
This is again expected from the effective actions \eqref{small q expansion of TT} and \eqref{large q expansion of TT}. 
The separated behavior in the region $s<0$ 
is due to the fact that the mid cycle $C_3$ is asymmetric with $C_1$ and $C_2$. 
This asymmetry appears in the sub-leading of the $q$-expansion in the small $q$ region, but in the large $q$ region it appears as a difference in the coefficient of leading of the $1/q$-expansion. 
This is the reason why the phase transition point splits at $s<0$.
Therefore, although it cannot be seen at the current resolution, a tiny separation of phase transition points in the small $q$ region is expected to occur as well as in the large $q$ region.

\section{Conclusion and Discussion}
\label{sec:Conclusion and Discussion}

In this paper, we have investigated the phase structure and duality of the fundamental Kazakov-Migdal (FKM) model on generic connected simple graphs.
The partition function of the FKM model is described by the graph zeta function weighted by unitary matrices on the links \cite{PhysRevD.108.054504}. 
In this paper, we have concentrated on the case where the graph zeta function becomes the Ihara zeta function $\zeta(q;U)$.
The series expansion of the Ihara zeta function guarantees that the effective action of the FKM model, $-\gamma N_c \log\zeta(q;U)$, can be expressed as a summation of all possible Wilson loops on the graph for small $q$. 
We found that the FKM model is unstable in the critical strip of the Ihara zeta function. 
We demonstrated that the FKM model on a regular graph has a small/large coupling duality associated with the functional relation of the Ihara zeta function. 
We also found a dual description of the Ihara zeta function on a general graph as a generalization of the function relation. 
Using this description, we have shown that the FKM model can be represented as the sum of all possible Wilson loops on the graph even in the region $q>1$.

As discussed in \cite{PhysRevD.108.054504}, the FKM model in large $N_c$ undergoes the GWW phase transition in general. 
When the fundamental cycles of minimal lengths are uniquely determined, the phase structure of the model at large $\gamma$ can be estimated from the results of the GWW model. However, in other cases, non-trivial coupling terms cannot be ignored.
We have checked the validity of this insight by using the numerical simulations and have shown that the system behaves as expected in the theoretical analysis. 
In particular, it is interesting that the behavior of the model in large $q$ region coincides with the theoretical prediction based on the dual description of the Ihara zeta function. 
The origin of this duality lies in the fact that the Ihara zeta function is represented through the sum of all possible Wilson loops.
This phenomenon is reminiscent of the duality of the string theory, 
and the FKM model is expected to provide a new perspective on the relationship between the string theory and the gauge theory.

While this paper primarily focuses on the Ihara zeta function, by incorporating more general mass parameters, it is possible to explore a broader class of graph zeta functions.
If we set the mass parameter $m_v$ as \eqref{eq:mv Bartholdi}, 
the partition function of the FKM model is expressed by the Bartholdi zeta function $\zeta_G(q,u)$ with the bump parameter $u$. 
It is straightforward to extend the discussion in this paper to this case,
where the phase diagram should be drawn in $(q,u)$-plane.
It will be discussed separately in future.

As we have seen in Fig.~\ref{Tetrahedron specific heat}, 
the behavior of the specific heat of the model on $K_4$ is different from that of the other graphs, that is,
the pointed ``elbow'' appears at the phase transition point.
As discussed already, 
it is reasonable to think that the effect of the interaction term is not making the order of the phase transition from third-order to second-order, 
but changing the behavior of the specific heat with keeping the order of the phase transition to be third-order. 
This observation also explains the small elbows that appeared in the model on graphs other than $K_4$ for finite $\gamma$
(see Figs.~\ref{C3 specific heat},\ref{Double triangle specific heat},\ref{Triangle-square specific heat} and \ref{Triple-Triangle specific heat}). 
Although the effective actions of these models reduce to the GWW model in large $\gamma$ where the elbow disappears and the phase transition is third-order, 
the interaction terms derived from the composite cycles remain finite for finite $\gamma$, 
which are expected to have a similar effect to the term $U_1U_2U_3$ in the $K_4$ case.
In order to make this point clear, we should consider the unitary matrix model with the action, 
\begin{equation}
  S = \frac{N}{\lambda}\Tr\Bigl[
  U_1 + U_2 + U_3 + g\, U_1U_2U_3 + h.c.
  \Bigr]\,,
\end{equation}
in close. 
It will also be discussed separately in another paper in preparation. 

\section*{Acknowledgments}
The authors would like to thank 
S.~Aoki, 
T.~Misumi 
and T.~Yoda
for helpful discussions.
This work is supported in part
by Grant-in-Aid for Scientific Research (KAKENHI) (C), Grant Number 20K03934 (S.~M.)
and Grant Number 23K03423 (K.~O.).


\appendix

\section{Poles of the Ihara zeta function of a regular graph}
\label{app:poles}

In this appendix, we show that all poles of the Ihara zeta function $\zeta_G(q)$ exist in the region $R_G\le q \le 1$ in general. 

Let us recall that the Ihara zeta function is expressed as $\zeta_G(q)=\det(1-qW)^{-1}$ where $W$ is the edge adjacency matrix \eqref{eq:matrix W}. 
Therefore, the poles of $\zeta_G(q)$ are given by the inverse of the eigenvalues of $W$. 
In particular, the maximal convergence radius $R_G$ is the inverse of the maximal eigenvalue of $W$. 

From this fact, we can immediately conclude the inequality \eqref{eq:range RG}. 
The point is that the matrix $W$ is a non-negative matrix. 
As a general property of a non-negative matrix, 
the maximal eigenvalue $\lambda_{\rm max}$ of $W$ satisfies 
\begin{equation}
  \min_{1\le \bse \le 2n_E}\left( \sum_{\bse'=1}^{2n_E}W_{\bse\bse'} \right)
  \le |\lambda_{\rm max}| \le 
  \max_{1\le \bse \le 2n_E}\left( \sum_{\bse'=1}^{2n_E}W_{\bse\bse'} \right) \,.
  \label{eq:general inequality}
\end{equation}
See e.g. \cite{horn2012matrix} for a proof. 
From the definition of the edge adjacency matrix \eqref{eq:matrix W}, $W$ satisfies 
\begin{equation}
  \sum_{\bse'=1}^{2n_E}W_{\bse\bse'} = \deg s(\bse) - 1 = t_{s(\bse)}\,. 
  \label{eq:property of W}
\end{equation}
Combining \eqref{eq:general inequality} and \eqref{eq:property of W}, 
we can conclude \eqref{eq:range RG} since $R_G = |\lambda_{\rm max}|^{-1}$.

To estimate the spectrum of $W$ in more detail, it is convenient to estimate $WW^T$; 
\begin{align}
  \sum_{\bsf\in E_D} W_{\bse\bsf}W_{\bse'\bsf}
  &= \sum_{\bsf\in E_D} 
  \left(\delta_{t(\bse)s(\bsf)-\delta_{\bse\bsf^{-1}}}\right)
  \left(\delta_{t(\bse')s(\bsf)-\delta_{\bse'\bsf^{-1}}}\right) \nn \\
  &= \left(\deg t(e) - 2\right) \delta_{t(\bse)t(\bse')} + \delta_{\bse\bse'}\,.
\end{align}
If we define 
\begin{equation}
  E_v \equiv \left\{\bse\in E_D | t(\bse)=v\right\}\,,
\end{equation}
and arrange $E_D$ as the disjoint union of $\{E_v| v\in V\}$, $WW^T$ can be expressed as a block diagonal matrix, 
\begin{equation}
  WW^T = \begin{pmatrix}
    F_{\deg v_1} & & \\
    & \ddots & \\
    && F_{\deg v_{n_V}}
  \end{pmatrix}\,,
\end{equation}
where $F_d$ is a matrix of size $d$ defined as 
\begin{equation}
  F_d \equiv (d-2)\begin{pmatrix}
    1 & \cdots & 1 \\
    \vdots & \ddots & \vdots \\
    1 & \cdots & 1 
  \end{pmatrix} + \bsone_{d} \,.
\end{equation}
Since it is easy to show 
\begin{equation}
  {\rm Spec}\, F_d = \left\{(d-1)^2,1,\cdots,1\right\}\,, 
\end{equation}
we see that
\begin{equation}
  {\rm Spec}\, WW^T = \left\{(\deg v_1-1)^2, \cdots, (\deg v_{n_V}-1)^2, 1,\cdots, 1\right\}\,. 
\end{equation}
In other words, the Rayleigh quotient of $WW^T$ with respect to a vector $\vec{v}$, 
\begin{equation}
  R(\vec v) \equiv \frac{\vec{v}\cdot WW^T\vec{v}}{|\vec{v}|^2}\,, 
\end{equation}
satisfies 
\begin{equation}
  1 \le R(\vec v) \le \left(\max_{v\in V}\deg v - 1\right)^2 = t_{\rm max}^2 \,. 
\end{equation}
If we choose $\vec{v}$ as an eigenvector of $W$ with the eigenvalue $\lambda$, the corresponding Rayleigh quotient is $R(\vec v)=\lambda^2$. 
Therefore, $\lambda$ satisfies 
\begin{equation}
  1\le |\lambda| \le t_{\rm max}\,, 
\end{equation}
and then the poles of the Ihara zeta function are in the region $R_G \le q \le 1$ 
since the maximal convergence radius is the norm of the minimal pole of $\zeta_G(q)$ by definition. 


\bibliographystyle{JHEP}
\bibliography{refs}

\end{document}